\documentclass{aa}  
\usepackage{natbib}         
\usepackage{graphicx}                    
\usepackage{amssymb}                    
\usepackage[usenames]{color}                         
\usepackage{txfonts}

\newcommand{\BE}{\begin{equation}}
\newcommand{\EE}{\end{equation}}
\newcommand{\BA}{\begin{eqnarray}}
\newcommand{\EA}{\end{eqnarray}}
 \newcommand{\fig}[1]{Fig.~\ref{fig_#1}}
 \newcommand{\figs}[2]{Figs.~\ref{fig_#1},\ref{fig_#2}}
 
 \newcommand{\sect}[1]{Sect.~\ref{sec_#1}}
 \newcommand{\sects}[2]{Sects.~\ref{sec_#1} and~\ref{sec_#2}}
 \newcommand{\eq}[1]{Eq.~(\ref{eq_#1})}
 \newcommand{\eqs}[2]{Eqs.~(\ref{eq_#1},\,\ref{eq_#2})}
 
\newcommand{\eg}{\textit{e.g.} }
\newcommand{\ie}{\textit{i.e.} }
\newcommand{\insitu}{\textit{in situ}}

\newcommand{\bmax}{b_{\rm max}}
\newcommand{\bmin}{b_{\rm min}}
\newcommand{\bmean}{b_{\rm mean}}
\newcommand{\Baxis}{B_{\rm axis}}
\newcommand{\Bx}{B_{\rm x}}
\newcommand{\By}{B_{\rm y}}
\newcommand{\Bz}{B_{\rm z}}
\newcommand{\BL}{\vec{B}_{\rm L}}
\newcommand{\BLi}{\vec{B}_{{\rm L},i}}
\newcommand{\Bmin}{r_{\rm Bmin}}
\newcommand{\Bo}{B_{0}}
\newcommand{\Bobs}{\vec{B}_{\rm obs}}
\newcommand{\Bobsi}{\vec{B}_{{\rm obs},i}}
\newcommand{\Bav}{<\!\!B\!\!>}
\newcommand{\Bxav}{<\!\!B_{\rm x}\!\!>}
\newcommand{\cha}{\mathcal{C}}
\newcommand{\dev}{{\it dev}}
\newcommand{\degree}{\ensuremath{^\circ}}
\newcommand{\dist}{{\it dist}}
\newcommand{\erf}{{\rm erf}}
\newcommand{\Np}{N_{\rm p}}
\newcommand{\pb}{\mathsf{P}}
\newcommand{\pbL}{\mathsf{P}_{\rm L}}
\newcommand{\pbG}{\mathsf{P}_{\rm G}}
\newcommand{\pp}{\mathcal{P}}
\newcommand{\pobsp}{\mathcal{P}_{\rm obs}}
\newcommand{\py}{\mathit{P}}
\newcommand{\rBxL}{r_{\rm Bx,L}}
\newcommand{\rBx}{r_{\rm Bx}}
\newcommand{\rotB}{\omega} 
\newcommand{\rotBmin}{\omega_{\rm min}}
\newcommand{\ysb}{y/b}

\newcommand{\grad}{ {\bf \nabla } }

\newcommand{\rmd}{ {\rm d } }

\begin{document} 

\title{Does the spacecraft trajectory strongly affect the detection of magnetic clouds?}

\author{P. D\'emoulin\inst{1} 
 \and    S. Dasso\inst{2,3}
 \and    M. Janvier\inst{1}
        }

\institute{
$^{1}$ Observatoire de Paris, LESIA, UMR 8109 (CNRS),
       F-92195 Meudon Principal Cedex, France \email{Pascal.Demoulin@obspm.fr}\\
$^{2}$ Departamento de F\'\i sica, Facultad de Ciencias Exactas y
Naturales, Universidad de Buenos Aires, 1428 Buenos Aires, Argentina \email{dasso@df.uba.ar}\\
$^{3}$ Instituto de Astronom\'\i a y F\'\i sica del Espacio, CONICET-UBA,
CC. 67, Suc. 28, 1428 Buenos Aires, Argentina \email{sdasso@iafe.uba.ar}\\
}

\date{Received ***; accepted ***}

   \abstract
   {Magnetic clouds (MCs) are a subset of interplanetary coronal mass ejections (ICMEs) where a  magnetic flux rope is detected. Is the difference between MCs and ICMEs without detected flux rope intrinsic or rather due to an observational bias?} 
   {As the spacecraft has no relationship with the MC trajectory, the frequency distribution of MCs versus the spacecraft distance to the MCs axis is expected to be approximately flat.  However, Lepping and Wu (2010) confirmed that it is a strongly decreasing function of the estimated impact parameter.  
Is a flux rope more frequently undetected for larger impact parameter?  }
   {In order to answer the questions above, we explore the parameter space of flux rope models,
especially the aspect ratio, boundary shape, and current distribution.  The proposed 
models are analyzed as MCs by fitting a circular linear force-free field to the magnetic field computed along simulated crossings. }
   { We find that the distribution of the twist within the flux rope, the non-detection due to too low field rotation angle or magnitude are only weakly affecting the expected frequency distribution of MCs versus impact parameter.  
However, the estimated impact parameter is increasingly biased to lower values as the flux-rope cross section is more elongated orthogonally to the crossing trajectory.  The observed distribution of MCs is a natural consequence of a flux-rope cross section flattened in average by a factor 2 to 3 depending on the magnetic twist profile.  
However, the faster MCs at 1 AU, with $V>550$ km/s, present an almost uniform distribution of MCs vs. impact parameter, which is consistent with round shaped flux ropes, in contrast with the slower ones.}
   {We conclude that either most of the non-MC ICMEs are encountered outside their 
flux rope or near the leg region, or they do not contain any.  }


   \maketitle


\section{Introduction} 
\label{sec_Introduction}

Interplanetary Coronal Mass Ejections (ICMEs) are detected in the solar wind (SW) by \insitu\ plasma and magnetic field measurements onboard spacecraft.  They are the counterpart of Coronal Mass Ejections (CMEs) observed with coronagraphs \citep[\eg ][]{Howard11,Lugaz11}.   With STEREO twin spacecraft having both \insitu\ and imager instruments, this link is presently well etablished \citep[\eg ][ and references therein]{Harrison09,Kilpua11,Rouillard11b,Lugaz12,Wood12}.   ICMEs are defined by one or several criteria \citep[see][ for reviews]{Wimmer-Schweingruber06s,Zurbuchen06}.   Typical criteria are: 
  (a) a stronger magnetic field with lower variance than in the surrounding SW; 
  (b) a low proton plasma $\beta_{p}$ ($<0.4$ typically); 
  (c) a smooth and large rotation of the magnetic field; 
  (d) a proton temperature at least lower by a factor 2 than in ambient SW with the same velocity as in the MC; 
  (e) an enhanced helium abundance (He/H $\geq 6$\%); 
  (f) the presence of counter-streaming suprathermal ($> 80$ eV) electron beams; 
  (g) enhanced ion charge states. 
Magnetic clouds (MCs) are defined with criteria (a-d) all satisfied \citep{Burlaga81}, then MCs are a sub-class of ICMEs. This signature has been modeled as a magnetic flux rope \citep[\eg ][]{Burlaga88, Lepping90,Lynch03,Dasso06,Leitner07}.

\citet{Gosling90} found that a MC is present inside ICMEs only for 10-30\% of the cases.      
Presently, in average, a MC is detected in about 30\% of ICMEs  \citep{Richardson10,Wu11}.  
\citet{Cane03} found that this ratio is evolving with the solar cycle: the MC/ICME ratio increases from $\approx 15\%$ at solar maximum to $\approx 100\%$ at solar minimum.   
They interpreted this evolution as due to an observational selection effect since CMEs are launched from higher solar latitudes at solar maximum than at minimum, so a spacecraft located in the ecliptic more frequently crosses the flux rope away from its axis at solar maximum 
than at solar minimum \citep{Richardson04}.  This evolution with solar cycle is confirmed by their newer results \citep{Richardson10} and by the results of \citet{Kilpua12} around the minimum between solar cycle 23/24 (they found that $\approx 76 \%$ of ICMEs have flux rope characteristics in the time period 2007-2010).  
Since nearly all ICMEs are MCs at solar minimum, it has been suggested that MCs are only observed when the spacecraft crosses the magnetic structure near the flux rope center 
\citep[\eg ][]{Jian06}. 

Still, the \insitu\ observations provide only a 1D cut through a 3D structure so there is 
a lack of information.
For MCs this is typically complemented by a fit of a magnetic model to the data, providing both the local orientation of the flux rope and its field distribution within the cross-section.  The most often used model, so far, is the so-called Lundquist model \citep{Lundquist50}, 
which considers a static and axi-symmetric linear force-free magnetic equilibrium configuration \citep[\eg\ ][]{Goldstein83,Burlaga88}. 
Its main advantage is its simplicity (low number of free parameters) while it satisfies the low plasma $\beta$ condition typically found in MCs (typically $\beta_{p}<0.1$) 
and fits relatively well observations \citep[e.g.][]{Lepping90,Burlaga95,Lepping03a,Dasso05}. 
The fit of the Lundquist model provides an estimation of the closest approach position (CA) of the spacecraft trajectory to the flux rope axis. 
It is generally expressed in \% of the flux rope radius $R$ \citep[e.g.,][]{Lepping10}. CA/$R$ is also called the impact parameter, and noted $p$ \citep[e.g.][]{Lynch03,Jian06}.  
The sign of $p$ indicates which side of the MC is crossed by the spacecraft. Below, we 
consider only the distance to axis, so $|p|$, and we simplify the notation to $p$.

\begin{figure}[t!]    
\centerline{\includegraphics[width=0.4\textwidth, clip=]{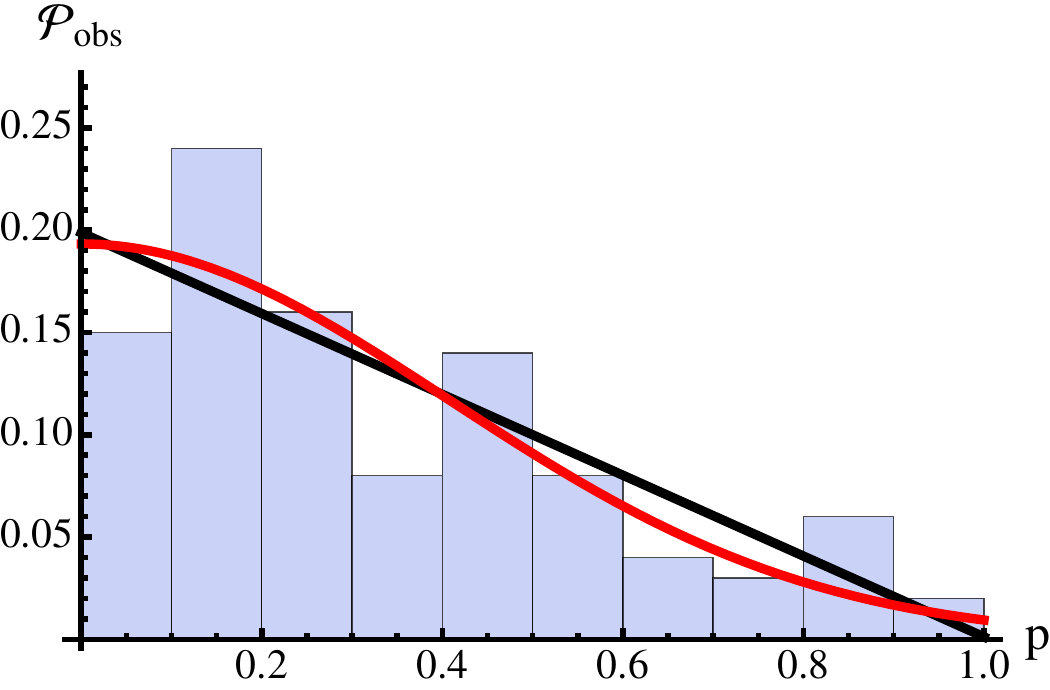}}
\centerline{\includegraphics[width=0.4\textwidth, clip=]{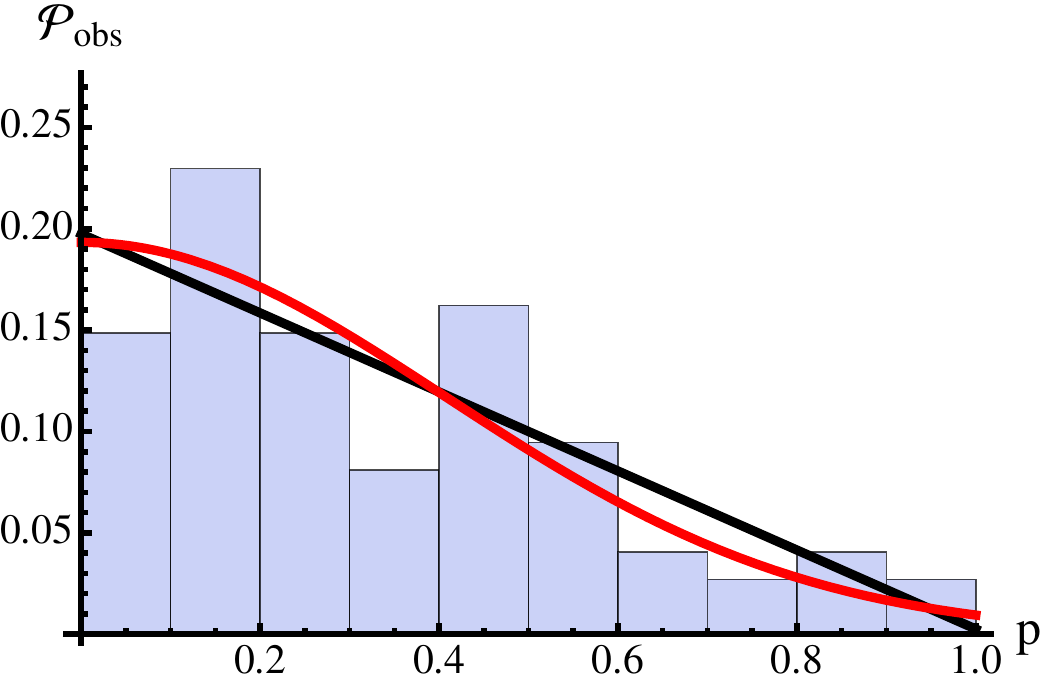}}
\caption{Probability distribution, $\pobsp (p)$, of the impact parameter ($p$). 
The results of MCs observed by WIND at 1~AU and fitted by the Lundquist model \citep{Lepping90,Lepping10}, are shown with a histogram having 10 bins of $p$.
Black curve: A linear fit to the histogram.
Red curve: The gaussian function derived by \citet{Lepping10}.
The histogram in the top panel has 100 MCs, while the one in the bottom panel is restricted to 
the 74 best observed MCs (quality 1,2). 
}
 \label{fig_Pobs_OKCA}
\end{figure}

With a set of 98 MCs observed at 1~AU, \citet{Lepping10} found that the number of MCs, detected at 1~AU nearby Earth, decreases with $|\textrm{CA}|$, or $p$ (\fig{Pobs_OKCA}, \sect{obs_Lepping}), confirming previous results \citep{Lepping06}.  They checked that the dependence with $p$ of the rotation angle and of the mean magnetic field along the spacecraft trajectory was behaving as expected for the Lundquist field model.  
While $p$ is the most uncertain parameter of the fit result for an individual MC \citep{Lepping90}, this self-consistency tests reinforce that, in average, $p$ was estimated correctly enough by the fit of the Lundquist model to the observations.   

Every MC with a flux-rope axis inclined on the ecliptic plane crosses it. Then,
because CMEs depart from the Sun at any longitude relative to the center disk, the related MC \insitu\ observations are expected to cross the flux rope at random distance from its axis.  Next, we consider  the minority of MC cases where the flux-rope axis is nearly parallel to the ecliptic plane. Because CMEs depart typically away from the solar equator, the spacecraft is expected to cross the flux rope at a distance to its axis which is correlated to its launched latitude.  Such cases imply a bias towards a larger impact parameter (even if the deflection of CMEs toward the heliospheric current sheet decreases this effect).  From these considerations, one expects a flat, or even slightly increasing, distribution of MCs versus $p$ which is not observed (\fig{Pobs_OKCA}).  

A first interpretation of the observed decrease (\fig{Pobs_OKCA}) is a strong selection effect due to a greater difficulty to detect a flux rope when $p$ is larger.  
In this case, correcting this selection effect by supposing a flat distribution, with the frequency detected for low $p$ value, would typically double the number of detected MCs.
Then, does a large fraction of the non-MC ICMEs correspond to undetected flux rope with the spacecraft trajectory being too far from the flux rope axis?
   
A weakness of the above analysis is that the deduced impact parameter can still be biased by the selection of a particular model. Indeed, the self-consistency tests of \citet{Lepping90} only check that the fit to the data provides coherent results with the hypothesis of the model.  Evidences of compression in the direction of propagation are present in CME observations \citep[e.g.][]{Savani09,Savani10} and in MHD simulations \citep[e.g.][]{Cargill02,Odstrcil04,Manchester04b,Lugaz05b,Xiong06}.  Such a compression flattens the cross section, and such geometrical feature has been partly taken into account by \citet{Vandas03}. They developed an extension of the Lundquist model from a circular to an elliptical boundary.  This model is still analytical (but relatively complex) and it introduces only one more parameter, the aspect ratio of the ellipse, if one supposes that the major axis of the elliptical cross section of the flux rope is perpendicular to the direction of its motion.  
Moreover, it provides a better fit to observed MCs having a relatively uniform field strength.  This flatness of the magnetic profile increases with the aspect ratio, pointing that some MCs have a relatively flat cross section \citep{Vandas05,Antoniadou08}. 
Finally, \citet{Vandas10} tested this model with the results of an MHD simulation by exploring several spacecraft crossings of the simulated flux rope.
They concluded that both the aspect ratio and the impact parameter $p$ are fully reliable only for low $p$ values, confirming and extending \citet{Lepping90} results. 

A variety of alternative models have been proposed for MCs.  Keeping the cylindrical symmetry, a variety of non-linear force-free field models are also possible (see \sect{cir_nlfff}). 
One possibility has a uniform twist within the cross-section \citep[e.g.][]{Farrugia99,Dasso03,Dasso05b}.  Also, several non force-free models have been applied, using different shapes for their cross sections \citep[e.g.][]{Mulligan99,Cid02,Hidalgo02,Hidalgo11}.  So far, even if a given model has been shown to better fit the data of a few MCs than other models, this conclusion has not been extended to a large set of MCs.  Indeed, the typical internal structure (\eg\ the twist profile) of MCs is still not precisely known.   
 
Another approach is to include the curvature of the flux rope axis by developing toroidal models 
\citep{Marubashi97,Romashets03,Marubashi07,Romashets09,Owens12}.  This approach is especially needed when a leg of the flux rope is crossed (\ie\ when the spacecraft trajectory is close to the local flux-rope axis direction).  In this case, the inclusion of the axis curvature can strongly change both the deduced axis orientation and impact parameter \citep{Marubashi12}.  Such leg crossings are typically characterized by a long duration MC with a complex rotation profile of the magnetic field as well as a low angle between the solar radial direction and the flux rope axis (called cone angle).  The frequency of such cases is small in the data set of \citet{Lepping10} with, for example only 6 over 98 MCs with a cone angle below $30\degree$, and none for the 67 MCs of quality 1,2 (as defined in their paper). Then, we consider only local models of MCs with straight axis as they have less free parameters than  toroidal models.

One major unknown is the extension of the MC cross section.  A way to deduce it is to solve the non-linear force free equations by a direct numerical integration with the measured vector magnetic field as boundary conditions.  This approach only supposes a magnetostatic field invariant by translation along the straight axis \citep{Hu02,Sonnerup06}.  The method was tested successfully with MCs crossed by two spacecraft \citep{Liu08,Kilpua09,Mostl09}.   The results depend on the MC studied, ranging from nearly round to elongated cross sections 
\citep{Hu05,Liu08,Mostl09,Mostl09b,Isavnin11,Farrugia11}.
The main limitation of this kind of approach is that it solves an ill-posed problem: the integration of an elliptic partial differential equation from a part of the boundary of the domain.  The results can indeed be strongly affected by the time resolution and range of the data used, as well as by the method used to stabilize the integration.
  
The short review above shows a large variety of flux-rope models. If the MC data of \citet{Lepping10} would have been fitted by one of the above flux rope models,
how would the estimated impact parameter, $p$, been affected?  Said differently, how model dependent is the MC distribution shown in \fig{Pobs_OKCA}?  Moreover, how strongly does selection effects, \eg\ on the amount of magnetic field rotation or field strength, affect such a distribution? We analyze these issues for a large set of MCs by studying a variety of force-free field models (as the plasma-$\beta$ is typically around $0.1$ in MCs).  The meaning of the main parameters used throughout the paper are summarized in Table~\ref{T-parameters}. 

The observation results and the fitting method are summarized in \sect{obs}.
In \sect{cir}, we investigate the effect of a broad range of magnetic field profiles ranging from flat to peaked around the axis, keeping a circular cross section. In \sect{elong} we mostly investigate the effect of the cross section elongation with models having elliptical cross sections.   We also analyze the effect of bending of the cross section to a ``bean shape''.   We conclude that $p$ is most affected by the aspect ratio of the cross section. Then, in \sect{aspect}, we deduce a distribution of the aspect ratio compatible with the results of \citet{Lepping10}.  Next, in \sect{subsets}, we further analyze the MCs according to their global properties and show that some set of MCs have a relatively round cross-section.  Finally, we conclude by summarizing our results and, in particular, answering the question set in the title of this paper (\sect{Conclusion}).

\begin{figure}[t!]    
\centerline{\includegraphics[width=0.5\textwidth, clip=]{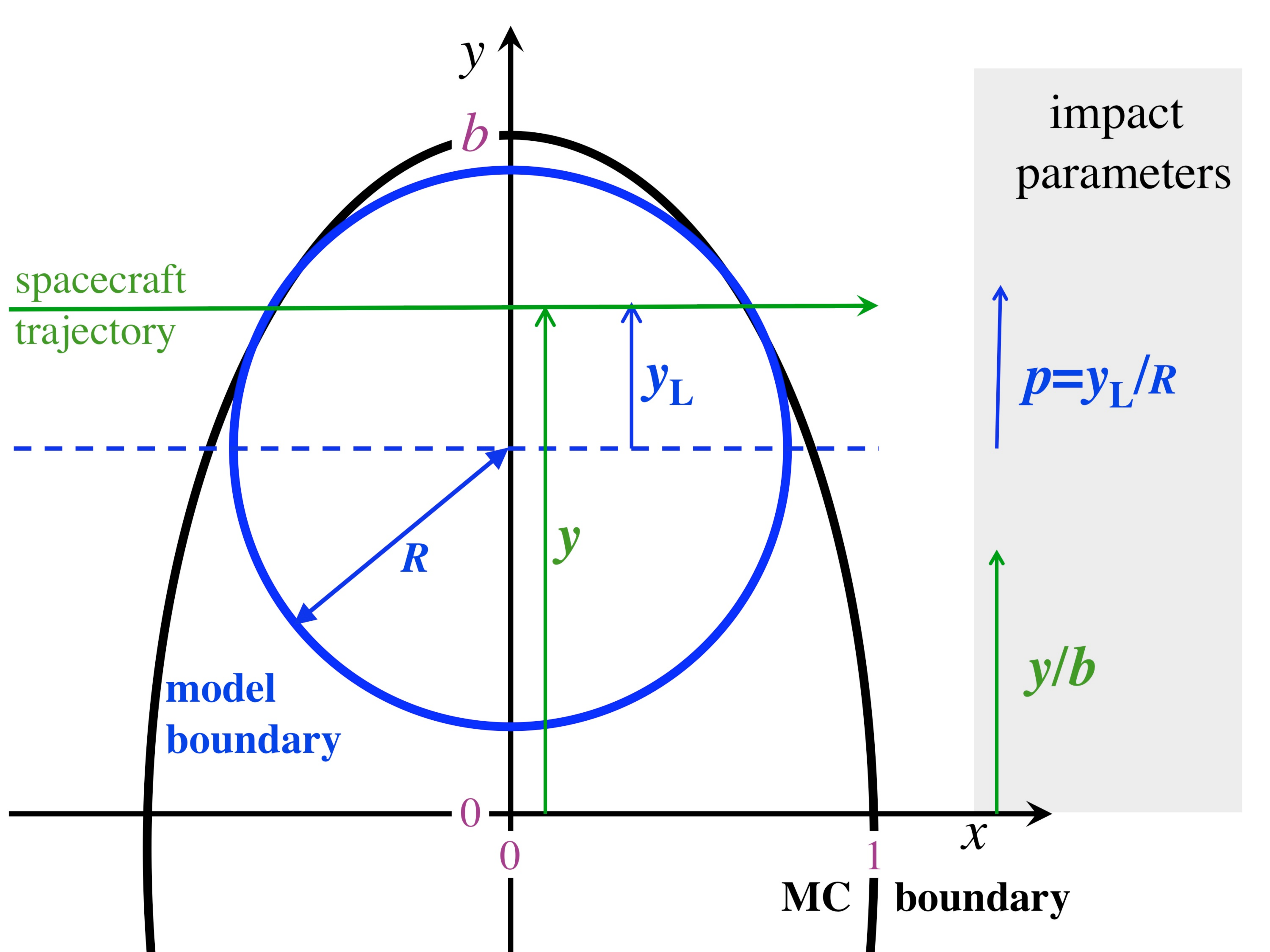}}
\caption{Drawing defining the geometry parameters for a spacecraft crossing a MC.
The fit of Lundquist field is schematized by the blue circle while the black ellipse 
delimitates the half extension of the MC boundary.  In this figure, we scale the drawing with 
the semi-minor axis of the ellipse set to unity. 
The true impact parameter, $y/b$, is larger than $p$.
}
 \label{fig_schema_p}
\end{figure}

\section{Observations and fitting method} 
\label{sec_obs}

\subsection{Observed probability distribution of the impact parameter} 
\label{sec_obs_Lepping}
We use the results of the Lundquist model fitted to MCs observed at 1~AU by WIND spacecraft from February 1995 to November 2007.  
They are available in Table~2 at http://wind.nasa.gov/mfi/mag\_cloud\_S1.html.
The list contains the results of 120 MCs at the date of 13 Dec, 2011. 
However removing the cases where the handedness could not be determined (flag f in the list) or the fitting convergence could not be achieved (flag F), this list restricts to 110 MCs. 
Next,  we examine the cone angle $\beta$ which is the angle between the MC axis (found by the Lundquist fit) to the solar radial direction (-{\bf X} axis in GSE coordinates).  We consider a folded angle, so $\beta$ is in the range $[0\degree,90\degree]$. Since the data obtained in the cases of MC leg crossing are the most difficult to analyze, making the fit results from cases with small $\beta$ angle the most uncertain (\sect{Introduction}), we limit the study to $\beta > 30 \degree$.  This restricts the MC sample to 103 MCs. Finally, there are 3 MCs with an impact parameter $p>1$ (so a fitted flux rope extending beyond the first zero of the axial field in the Lundquist model).
Removing these suspicious cases, all of the worse class  \citep[quality 3, as defined in][]{Lepping10}, it remains 100 MCs.  One can even be more strict on the selection criteria.   An extreme case is to select only the best cases (quality 1), with the limitation that the statistics is restricted to 19 MCs.  A less extreme case is to select the best and good cases (quality 1 and 2) so 67 MCs.  We verify that our results are not significantly affected by the group of MC selected.   

\citet{Lepping10} found that the number of detected MCs decreases rapidly with $p$.  The same result is shown in \fig{Pobs_OKCA} for the 100 MCs selected and a bin size of $\Delta p =0.1$.  Very close results are obtained if we restrict the analysis to the best observed MCs (quality 1 and 2).
  We define a probability of detection by normalizing the sum of the bin counts to 1.  This allows the comparison between the results obtained with different sets of MCs and the model predictions. The Gaussian function shown in \fig{Pobs_OKCA} (red curve) is a fit of the distribution as given by Eq.~(A1) of \citeauthor{Lepping10} with $b=0$ and $\sigma=0.407$. It also fits well both sets of MCs shown in \fig{Pobs_OKCA}.  The observed distribution can also be fitted with a linear function without significant difference with a Gaussian function (taking into account the statistical fluctuations).  
   
\begin{figure*}[t!]    
\centerline{\includegraphics[width=\textwidth, clip=]{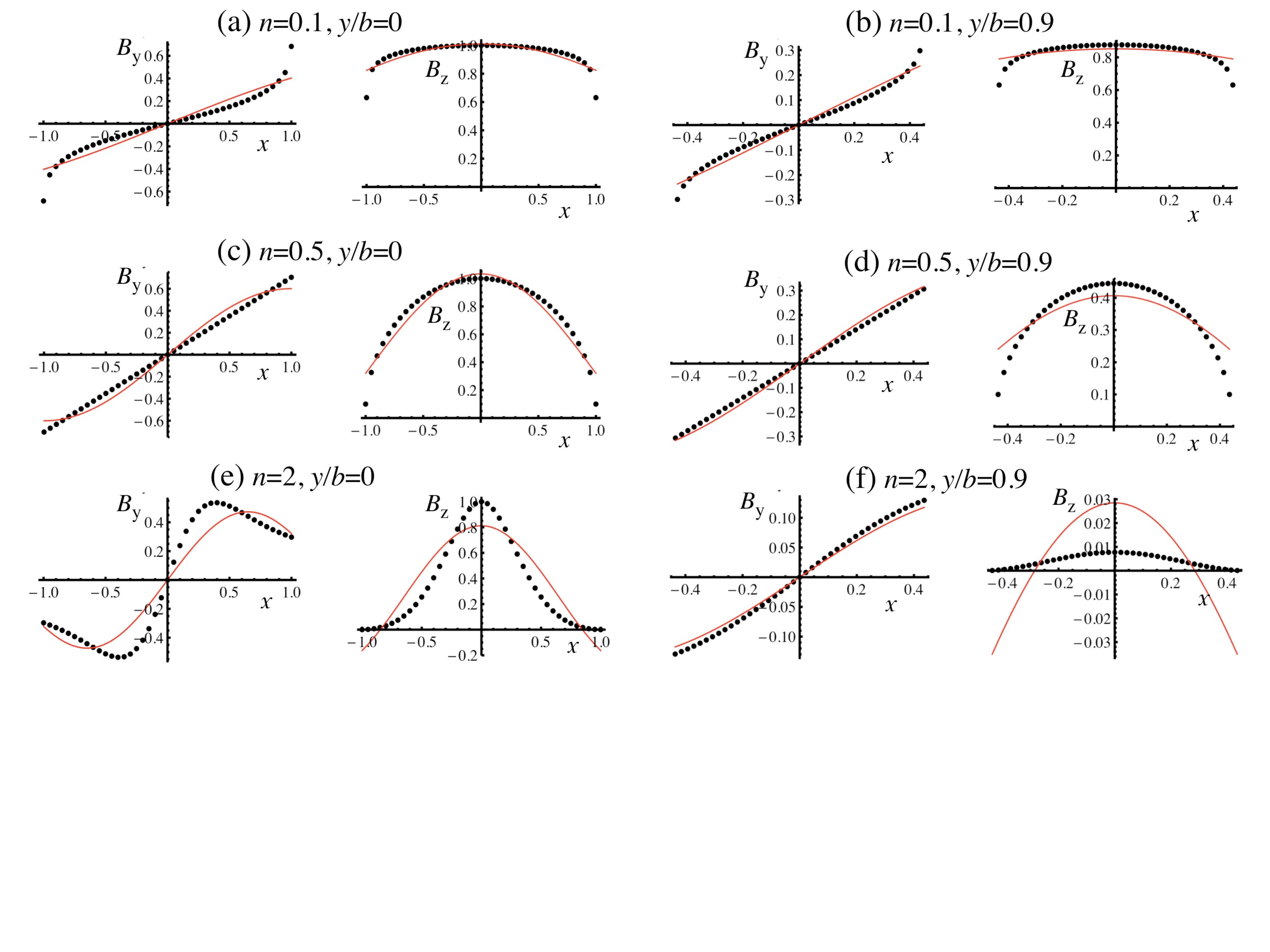}}
     \vspace{-0.22\textwidth}   
\caption{Examples of circular models (black dots) least square fitted with the Lundquist field (red curves).  Three non-linear force-free models ($n=0.1,0.5,2$) are selected to represent strong departure to the Lundquist field ($n=1$).  The true impact parameter, $\ysb$, is either null (left) or large (right).  $\Bz$ is the axial field component, and $\By$ is the azimuthal field component for $y=0$.  The field strength of the model on the axis, $\Baxis$, is normalized to 1.
}
 \label{fig_BfitsAn}
\end{figure*}

\subsection{Flux rope fit with the Lunquist field} 
\label{sec_obs_fit}

  Models are used to simulate flux rope crossings, providing synthetic observations which are analyzed as MCs, so following the classical procedure of \citet{Lepping90}. Those synthetic models can be chosen as circular or elliptic, and the half extension of such a structure
is given in black in \fig{schema_p}.  Then, the bias in the fits are analyzed.
The simulated trajectory is set at a distance $y$ parallel to the $x$-axis (because of the invariance in the $z$ direction of the models, the same $\vec{B}$ would be obtained along a trajectory inclined on the flux rope axis).  The true impact parameter is $\ysb$ where $b$ is the size of the structure in the $y$-direction (\fig{schema_p}).
The synthetic observations are fitted with a linear force-free model, the classical Lundquist solution \citep{Lundquist50}, which is in cylindrical coordinates:  
  \BE \label{eq_Lundquist}
   \BL = \Bo ~(~0,J_1(\alpha r),J_0(\alpha r)~) \,,  
  \EE
where $\alpha$ is associated with the first zero of $Bz$, 
$\Bo$ is the axial field strength and $J_m$ is the ordinary Bessel function of order $m$.  The fit of \eq{Lundquist} to the synthetic observations provides an estimation of $y$, called $y_{\rm L}$, with an origin not necessarily located on the true flux rope axis (\fig{schema_p}). Then, the Lundquist fit provides the estimated impact parameter $p=y_{\rm L}/R$ where $R$ is the estimated flux rope radius (defined for $\Bz =0$). 

The $\BL$ field is fitted to the synthetic observations $\Bobs$ by minimizing the function $\dev$ defined by:
  \BE \label{eq_dev}
   \dev = \sqrt{ \frac{1}{\Np} \sum_{i=1}^{\Np} (\BLi-\Bobsi)^2 } \,,  
  \EE
where $\Np$ is the number of points in the synthetic observations, and $\Bobs$ is related with $y/b$, while $\BL$ is related to $p$.  Providing that $\Np$ is large enough
(\ie $\Np >20$), the results of the fits are insensitive to the value of $\Np$, as expected since the synthetic observations are well resolved with such $\Np$ values.  Since the orientation of $\Bobs$ in MCs is following that of $\BL$ better than the magnetic field magnitude, \citet{Lepping90} fitted $\Bobs$ with $\BL$ with a two steps procedure.  In a first step, both $\Bobs$ and $\BL$ norms are normalized to unit at each point before minimizing $\dev$.  
Then, in a second step, the full fields are considered and $\dev$ is minimized by only changing the axial field strength $\Bo$.
From synthetic Lundquist fields, \citet{Gulisano07} also concluded that fitting to normalized $\Bobs$ gives better estimation of the real orientation of the MC axis. 

Compared to real MC observations, the models provide synthetic observations with no internal structures, and with known axis orientation and boundaries.   The exploration of the effect of perturbations to $\vec{B}$, various axis orientation and boundaries could be realized in the line of the following  exploration of the parameter space (\eg , the parameters defining the shape of the cross section).  However, we choose to rather limit the exploration to the global structure of the flux ropes (\ie the magnetic field repartition and the cross section shape) as such structure is expected to have a main effect on the estimated impact parameter.  Then, in the above first step, the minimization is realized by changing $p$ and $\alpha$ 
(as both the axis orientation and the flux rope boundaries, set at $\Bz =0$, are known and fixed).

\section{Detecting circular flux ropes} 
\label{sec_cir}

In this section, we analyze a series of circular force-free fields in order to test if the impact parameter could be biased by the choice of the model with the classical analysis of \citet{Lepping90}. 

\subsection{Force-free models} 
\label{sec_cir_nlfff}

Frequently, the magnetic structure of a MC is locally approximated by a straight flux rope invariant along its axis (\sect{Introduction}).  We use below an orthogonal frame, called the MC frame, with coordinates $(x,y,z)$. The $z$ direction is along the MC axis.  
$\vec{B}$ being independent of $z$ and $\grad \cdot \vec{B}=0$ imply together that one can write the magnetic field components orthogonal to the symmetry axis as: 
   $B_x = \partial A/ \partial y$ and $B_y = -\partial A/ \partial x$, 
where $A(x,y)$ is the magnetic flux function.   The force-free field condition implies
  \BE \label{eq_nlfff}
  \triangle A + \frac{\rmd B_z^{2}/2}{\rmd A} = 0 
        ~{\rm , ~~~~with~} B_z(A) \,.  
  \EE

A series of non-linear force-free fields are generated by      
  \BE \label{eq_B_z(A)}
   B_z(A) = c ~A^{n} \,,  
  \EE
where $c$ and $n>0$ are independent of $x,y,z$.  Typically, the flux rope boundary is set at a location where $B_z=0$ which can be set for $A=0$ without loosing generality. 
We normalize the cross section extension to the half of its maximal value in the $x$-direction (\fig{schema_p}).  Its half maximal extension in the $y$-direction is then the aspect ratio, $b$, set to $b=1$ in this section (i.e., circular shape). The flux rope axial field is called $\Baxis $.  

  Since we consider circular flux ropes in this section, \eq{nlfff} reduces to a differential equation of second order with the radius ($\sqrt{x^2+y^2}$). It is solved by a numerical integration using a shooting method \citep[\eg][ p.746]{numerical-recipes} applied to the resonance problem set by \eqs{nlfff}{B_z(A)} and the three boundary conditions: $A(0)=1$, $[\rmd A/\rmd r](0)=0$ and $A(1)=0$ (corresponding respectively to an azimuthal flux normalized to $1$, a regular field on the axis, and to $B_z =0$ at the boundary). We select the lowest eigenvalue $c$ to have models with an axial field vanishing only at the boundary as present in most MCs. For $n=1$, the field is a linear force-free field and $c=\alpha$ (as defined by \eq{Lundquist}). Finally the field strength on the axis can be scaled to any desired $\Baxis$ value.

The axial electric current density is:
  \BE \label{eq_jz}
   j_z = -\triangle A / \mu_0 = n~c^2 ~(A(x,y))^{2n-1} / \mu_0 \,,  
  \EE
where $\mu_0$ is the permeability of the free space.
For $n=0.5$, $j_z$ is uniform, while for $n>0.5$, $j_z$ decreases from the axis to the boundary (where $A=0$, so $j_z=0$). Finally, for $n<0.5$, $j_z$ is singular at the boundary (presence of a current sheet).
    
The value of $n$ also determines the spatial variation of $\vec{B}$.   This is illustrated in the left panels of \fig{BfitsAn}.    As $n$ is increased, the magnetic field strength becomes more concentrated around the axis and, near the boundary for $n>0.5$, the azimuthal field ($=|\By |$ at $y=0$) is a decreasing function of the radius ($=\sqrt{x^2+y^2}$) over a larger radius range. 
The case $n=2$, \fig{BfitsAn}e, is an extreme case for a MC.   At the opposite, as $n$ decreases the profile of the magnetic field strength flattens. For $n=0.5$, the azimuthal field is linear with radius, while for $n<0.5$ it is increasing more sharply near the boundary as $n$ is decreased.  The case $n=0.1$ is another extreme case for a MC.   The Lundquist model fits relatively well the different models except in extreme cases (\eg\ $n=2$ case, \fig{BfitsAn}e,f), even for large impact parameters (\eg\ see the case $y/b=0.9$ in \fig{BfitsAn}b,d).

\begin{figure}[t!]    
\centerline{\includegraphics[width=0.5\textwidth, clip=]{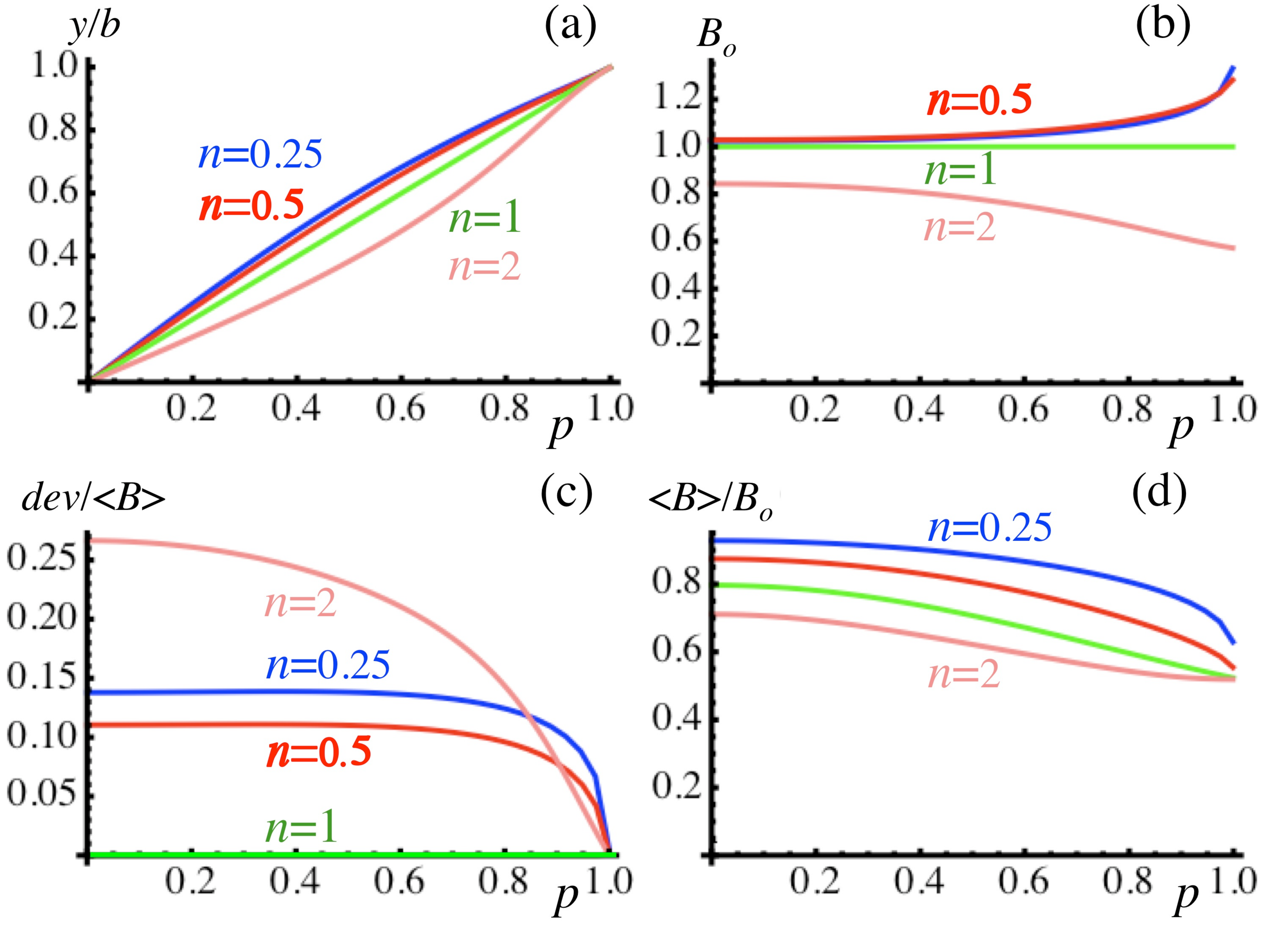}}
     \vspace{-0.02\textwidth}   
\caption{{\bf (a)} Dependence of the true impact parameter, $\ysb$, {\bf (b)} the fitted Lundquist field strength on the axis, $\Bo$,  {\bf (c)} the normalized deviation, $\dev /\!\!\!\Bav$, and {\bf (d)} the mean field magnitude $\Bav$ normalized to $\Bo$ in function of the impact parameter, $p$, found by fitting the Lundquist field to the models.  The parameter $n$ describes the profile of the axial component of the magnetic field and electric current, see \eqs{B_z(A)}{jz}.  
}
 \label{fig_resAn(p)}
\end{figure}

\begin{table} 
\caption{Description of the main parameters and where they are defined.}
\label{T-parameters}
  {\small 
\begin{tabular}{l@{~~}l}     
  \hline 
\multicolumn{2}{l}{\bf \qquad Synthetic models} \\
  $a$      & bending of the boundary                     \hfill \sect{elong_bent}\\  
  $b$      & aspect ratio of the boundary                \hfill \fig{schema_p} \\
  $\Baxis$ & field strength on the flux rope axis        \hfill \sect{cir_nlfff}\\
  $c$      & $\Bz$ for $A=1$,                            \hfill \eq{B_z(A)} \\
  $\cha$   & $=\{a,b,n,\Bmin,\rotBmin \}$, set of all characteristics \\
           & \quad of a model and selection parameters   \hfill \eq{P(p)} \\
  $n$      & exponent defining $\Bz$ and $j_{\rm z}$     \hfill \eqs{B_z(A)}{jz} \\  
  $x$      & coordinate along the simulated trajectory   \hfill \fig{schema_p} \\
  $y$      & coordinate across the simulated trajectory  \hfill \fig{schema_p} \\
  $z$      & coordinate along the flux rope axis         \hfill \sect{cir_nlfff} \\
  $\ysb$   & true impact parameter                       \hfill \fig{schema_p} \\
            
\multicolumn{2}{l}{\bf \qquad Fitted Lundquist model} \\           
  $\alpha$        & linear force-free field constant     \hfill \eq{Lundquist} \\
  $\BL$           & Lundquist field                      \hfill \eq{Lundquist} \\
  $\Bo$           & estimated axial field strength       \hfill \eq{Lundquist} \\
  $\dev$          & function of fit minimization         \hfill \eq{dev}  \\
  $R$             & flux rope radius (for $\Bz =0$)      \hfill \fig{schema_p}\\
  $y_{\rm L}$     & estimated distance of the spacecraft trajectory \hfill \fig{schema_p} \\
                  & \quad to the flux rope axis          \hfill \\
  $p$             & $=y_{\rm L}/R$, estimated impact parameter \hfill \fig{schema_p} \\
  
\multicolumn{2}{l}{\bf \qquad Estimated along the spacecraft trajectory} \\           
  $\Bav$          & average $\vec{B}$ strength           \hfill \sect{cir_info} \\
  $\Bxav$         & average $\vec{B}$ component parallel to the spacecraft  \hfill \sect{elong_aspect_ratio} \\
                  & \quad trajectory                     \hfill \\
  $\rBx$          & $=\; \Bxav \!\! / \!\!\!\Bav $       \hfill \sect{elong_aspect_ratio} \\                   
  $\rotB$         & rotation angle of $\vec{B}$ across the flux rope  \hfill \sect{cir_selection} \\ 
  
\multicolumn{2}{l}{\bf \qquad Selection parameters} \\           
  $\Bmin$         & minimum average field strength to detect   \\  
                  & \quad a flux rope: $\Bav \! / \Baxis \geq \Bmin $  \hfill \sect{elong_Expected}\\
  $\rotBmin$      & minimum rotation angle of $\vec{B}$ to detect a flux rope ~  
                      \hfill \sect{elong_Expected}\\

\multicolumn{2}{l}{\bf \qquad Probability functions} \\           
  $\pobsp (p)$    & observed probability                                  \hfill \fig{Pobs_OKCA} \\  
  $\pp (p,\cha)$  & theoretical probability for a model defined by $\cha$ \hfill \eq{P(p)} \\  
  $\pp (p)$       & $\pp (p,\cha)$ integrated on a flux-rope set          \hfill \eq{P(p)int} \\  
  $\pb (b)$       & probability distribution of $b$                       \hfill \eq{P(p)int}\\
  $\bmean$        & mean value of $b$                                     \hfill \eqs{bmean}{pbG}\\
  \hline
\end{tabular}
     }  
\end{table} 

\subsection{Informations provided by the Lundquist fit} 
\label{sec_cir_info}
 
The results of \fig{resAn(p)}a show that $\ysb$ is relatively well estimated by $p$ for all the range of $n$ values relevant to MCs.  The extreme case $n=0.1$ has very similar results to the case $n=0.25$, so it is not shown.  All cases with $n<1$ have $p$ only slightly lower than $\ysb$ so that the observed distribution probability (in function of $p$) would only be slightly compressed towards lower $p$ 
values compared to the original distribution probability (in function of $\ysb$).  On the contrary, all cases with $n>1$ would introduce a bias opposite to those observed since $p > \ysb$ (\fig{Pobs_OKCA}).  We conclude that the deviation around $\ysb =p$ (green curve, $n=1$ in \fig{resAn(p)}a) cannot explain the strong decrease of the probability to observe a MC with moderate and large $p$ values seen in \fig{Pobs_OKCA}.

The axial field of the model, set to $\Baxis =1$, is also well recovered with the fitted parameter $\Bo$ of the Lundquist field (\fig{resAn(p)}b).  Only large differences ($>20$\%) are obtained for glancing encounters ($p$, so $\ysb$, close to 1) or peaked magnetic field profiles (\eg $n=2$).  

Next, we normalize the deviation, \eq{dev}, by $\Bav$, the average of the field strength along the simulated trajectory.  Figure~\ref{fig_resAn(p)}c shows that the deviation of the fit to synthetic data is relatively small unless extreme cases are considered (\eg $n=2$, see also \fig{BfitsAn}). It is also important to notice that $\dev /\!\!\Bav$ is not a secure indicator of the precision of the fitted parameters: for example $\BL$ fits well the synthetic observations for large $p$ with a low $\dev /\!\!\Bav$ value (\figs{BfitsAn}{resAn(p)}) while the fitted parameters are more biased (\eg\ $\Bo$) and/or unprecised for these glancing encounters.

Finally, \citet{Lepping10} found that $\Bav \!\!/\Bo(p)$ from the analyzed MCs was following very well the expected relation from the Lundquist field except from a slight shift in ordinate (see their Fig.~5).  This shift could be explained by $n\approx 0.7$--$0.8$ (\fig{resAn(p)}d). This is an indication that the typical field in MCs is in between a linear force-free field and a field with constant axial current density. This result is in the same line of those obtained by \citet{Gulisano05}, where crude approximations were done (as assuming circular cross section, zero impact parameter, and the orientation of the main axis given only from the minimum variance method). 
They found clues in favor of magnetic configurations 
in between linear force-free field and constant current models.

\begin{figure}[t!]    
\centerline{\includegraphics[width=0.4\textwidth, clip=]{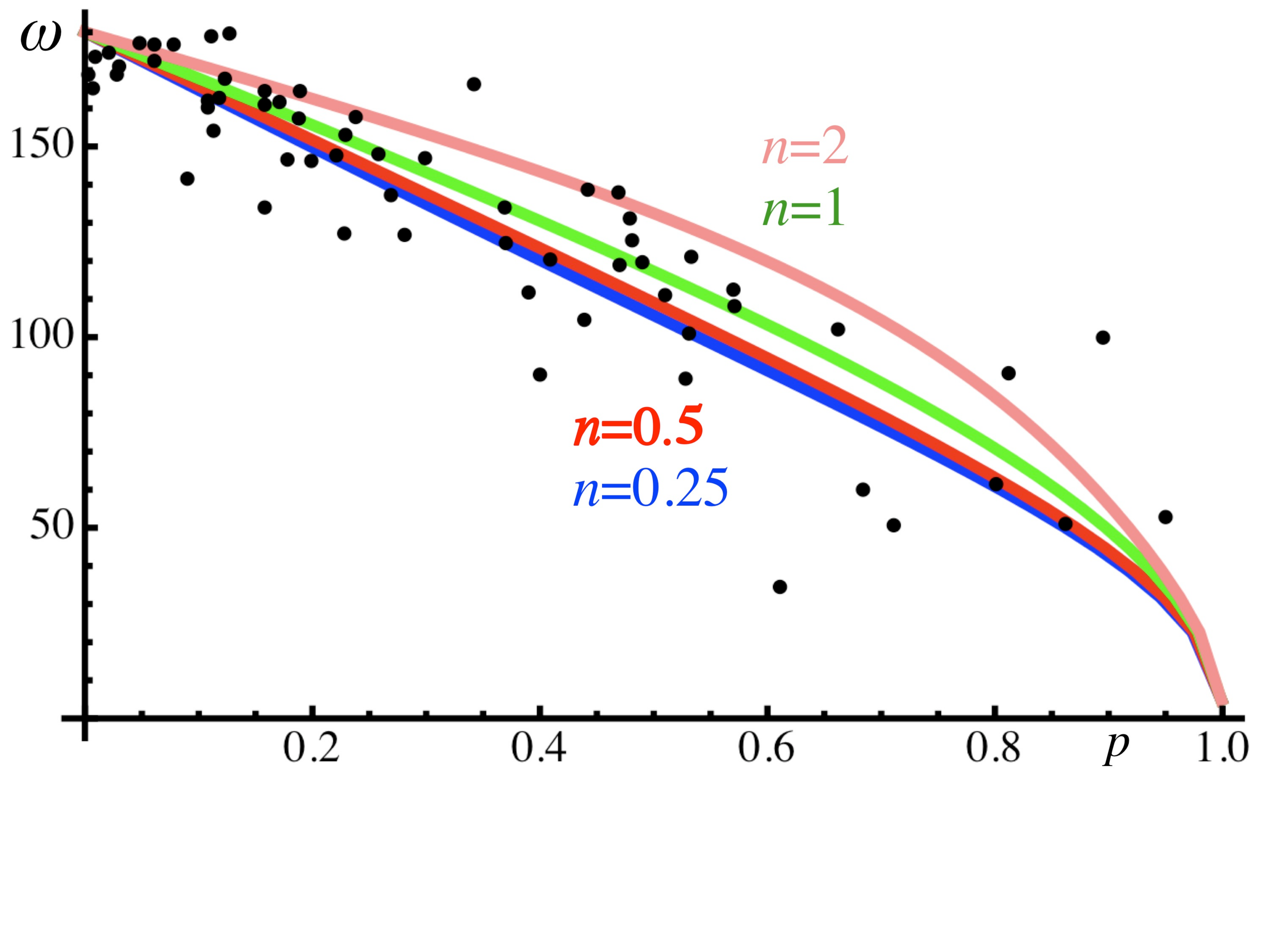}}
     \vspace{-0.06\textwidth}   
\caption{Dependence of the rotation angle, $\rotB$, of the magnetic field component orthogonal to the axis in function of the impact parameter, $p$, found by fitting the Lundquist field to circular models (\eqs{nlfff}{B_z(A)}).  The dots are the results obtained by \citet{Lepping10} for 65 MCs (see their Fig. 3B).   
}
 \label{fig_rotAn(p)}
\end{figure}

\subsection{Is there a significant selection effect with $p$?} 
\label{sec_cir_selection}
 
The rotation angle of the magnetic field along the simulated trajectory is weaker as the impact parameter increases (\fig{rotAn(p)}).  Since an important field rotation is one key ingredient to define a MC, a too weak rotation angle could lead to no MC detection, so a bias in the probability distribution in function of $p$. As in \citet{Lepping10}, we analyze the rotation angle of $\vec{B}$, in the plane orthogonal to the flux rope axis, by taking the angle formed by $\vec{B}$ at each of the two boundaries (hereafter noted $\rotB $).  As they averaged the observed $\vec{B}$ over 1~hour (their Fig.~3B), and to be comparable to the observations, the modeled $\vec{B}$ are averaged over 5\% of the crossing length near the boundaries.  For a typical MC duration of 20h, this implies an average over 1h, so that our results are directly comparable to their Fig.~3B. 

  For the case $n=1$ the synthetic data are derived from the Lundquist field, so as the fitting field; it implies that the rotation angle has a simple expression, $2 \arccos (p)$, which is nearly identical to the green curve in \fig{rotAn(p)} (the differences are only due to the small averaging performed near the boundaries).  Finally, we found that a broad range of $n$ values are compatible with their results (\fig{rotAn(p)}) so that the amount of $\vec{B}$ rotation angle is not selective of different models.  
This result is confirmed in \sects{elong}{aspect}.

Is there a severe selection effect with the amount of rotation angle in observed MCs?  
In fact, for a rotation angle lower than $30\degree$ no MC is observed (\fig{rotAn(p)}). Four MCs are observed with a rotation angle as low as $\approx 50\degree$, or lower, showing that MCs with low rotation angle can be detected.  This value of $\rotB$ corresponds to $p>0.9$ with the above models, so that a selection effect on MCs with low $\vec{B}$ rotation angle cannot explain the progressive decrease of the detection probability of MCs with $p$ (\fig{Pobs_OKCA}).  A way out would be to argue that significant $\vec{B}$ structures are frequently present within MCs, especially close to the boundary (\ie for large $p$ values), so that they can mask the lower rotation cases even more.  However, an important rotation angle, $\rotB > 90\degree$, is still present for $p \approx 0.7$ both for MCs and simulated flux ropes (\fig{rotAn(p)}).   Then, it is very unlikely that the presence of $\vec{B}$ structures in MCs can decrease the probability of detection by a factor 3 to 4 for $p \approx 0.7$ (\fig{Pobs_OKCA}).  It implies that a selection effect on rotation angle cannot explain the observed probability.  In the same line, $\Bo$ remains close to $\Baxis$ (\fig{resAn(p)}b) so that a selection effect on the field strength is expected to be weak for flux rope with nearly circular cross section.

\textbf{We conclude that the explored circular models cannot explain the observations} (\fig{Pobs_OKCA}).

\begin{figure}[t!]    
\centerline{\includegraphics[width=0.5\textwidth, clip=]{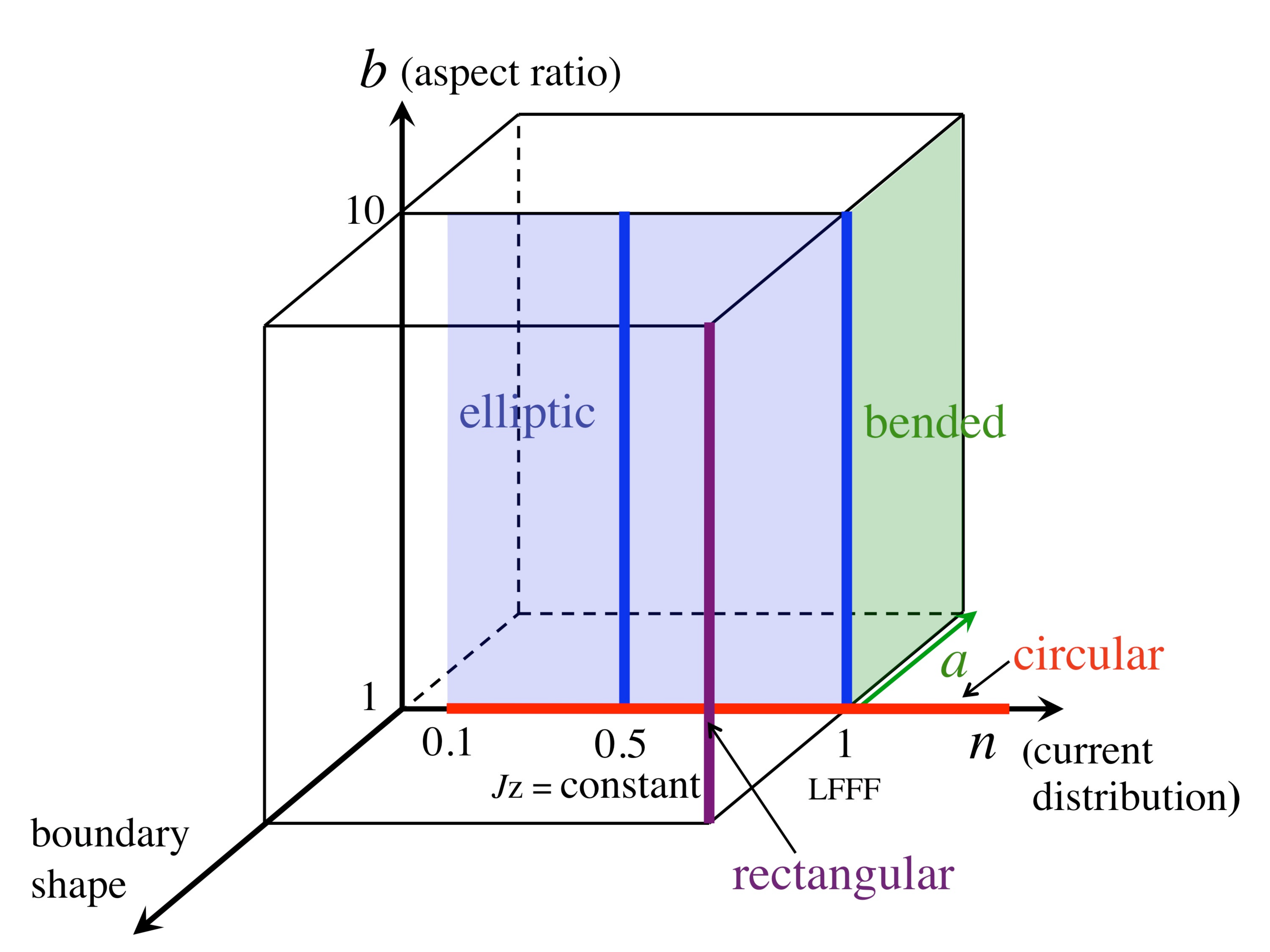}}
\caption{Drawing defining the regions of the parameter space explored. The red line indicates the circular models analyzed in \sect{cir}.  The MC boundary is elliptical for the blue region and it is deformed to a bean shape in the green region.  The two blue lines indicate the elliptical models analyzed in \sect{elong}.  Finally the purple line indicates an extreme case where the MC boundary is rectangular. $n$ defines the axial electric current and magnetic field component (\eqs{nlfff}{B_z(A)}). A cross section elongated orthogonally to the spacecraft trajectory has $b>1$ (\fig{schema_p}).
}
 \label{fig_schema_cube}
\end{figure}

\begin{figure}[t!]    
\centerline{\includegraphics[width=0.5\textwidth, clip=]{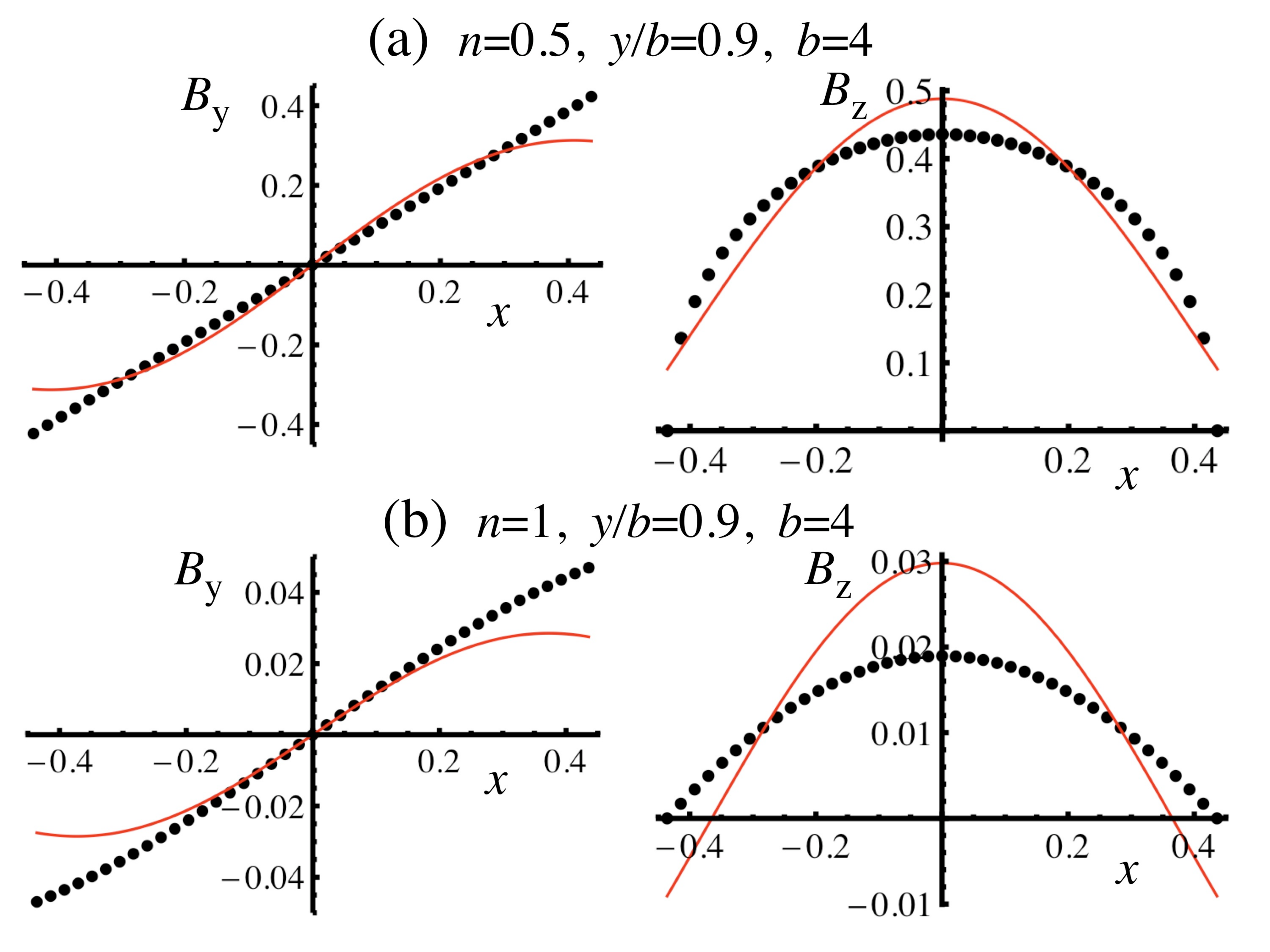}}
\caption{Examples of two elliptical models ($n=0.5,1$, black dots) least square fitted with the Lundquist field (red curves) for a large true impact parameter, $\ysb=0.9$.  $\Bz$ is the axial field component, and $\By$ is the field component both orthogonal to the simulated trajectory and to the flux rope axis.  
}
 \label{fig_BfitsEJ}
\end{figure}

\begin{figure}[t!]    
\centerline{\includegraphics[width=0.5\textwidth, clip=]{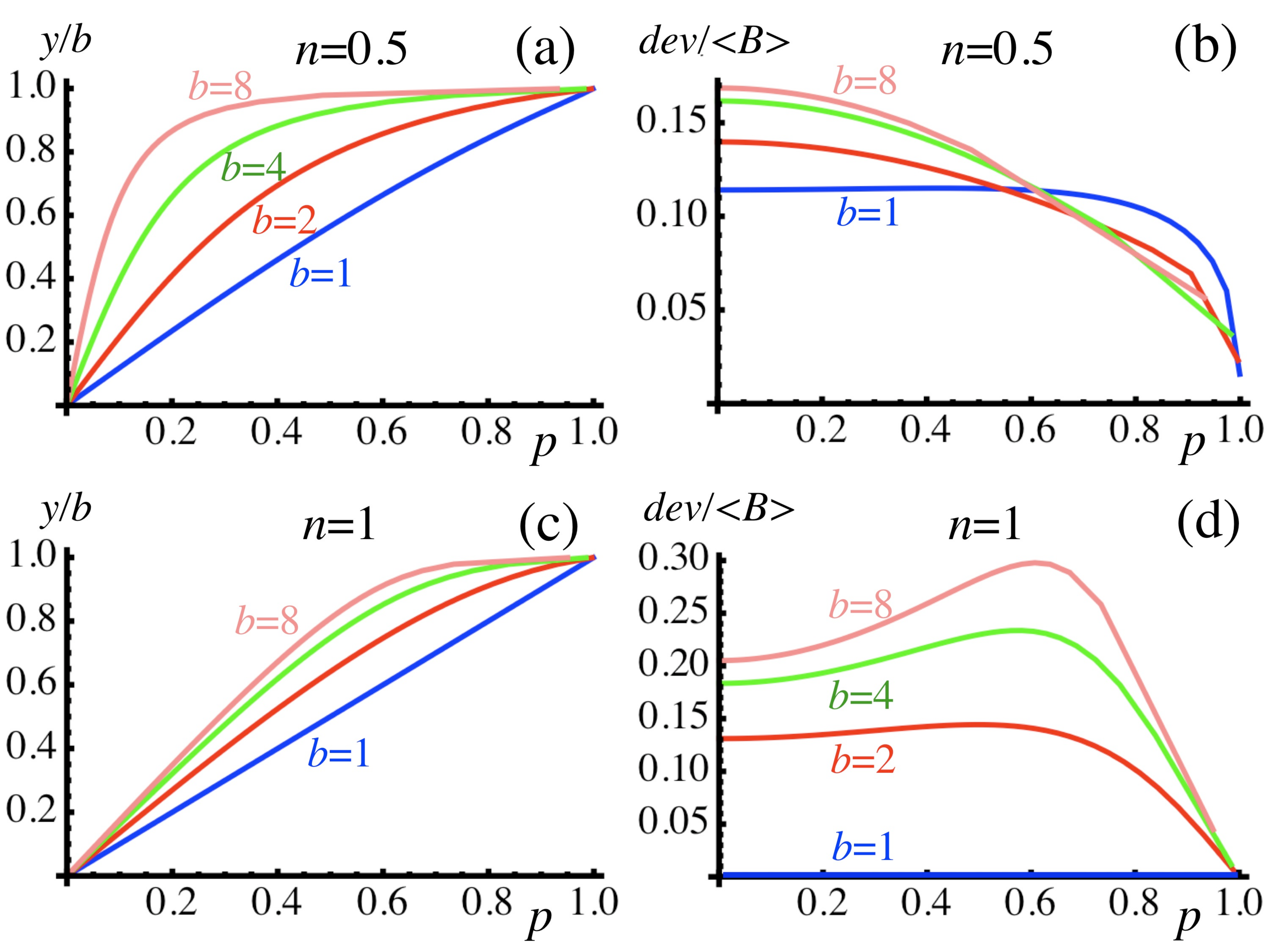}}
\caption{Dependence of the true impact parameter, $\ysb$, and of the normalized deviation, $\dev /\!\!\!\Bav$ in function of the fitted impact parameter, $p$, found by fitting the Lundquist field to elliptical boundary models.  The parameter $n$ describes the profile of the axial current (\eqs{nlfff}{B_z(A)}).  
}
 \label{fig_resEJ(p)}
\end{figure}

\section{Detecting flux ropes with elongated cross-section} 
\label{sec_elong}

In this section, we explore mainly the effects of the flux-rope boundary shape. An elliptical boundary is described by its aspect ratio $b$ ($b>1$ means that the cross section is elongated orthogonally to the MC trajectory, \fig{schema_p}). We also consider bent cross sections in \sect{elong_bent} described by an extra parameter called $a$.  A rectangular cross section is also considered as an extreme case.  These different types of cross sections allow to explore the space of parameters (\fig{schema_cube}) with models depending on a set of parameters ($a,b,n$).

\subsection{Expected effect of the cross-section aspect ratio} 
\label{sec_elong_aspect_ratio}

 \citet{Gulisano07} have shown that the ratio $\rBx\!\! =\; \Bxav \!\! / \!\!\!\Bav $ is a function of the true impact parameter $\ysb$ for a variety of circular models ($<>$ means averaging along the spacecraft trajectory within the MC).
\citet{Demoulin09b} have extended this relationship for linear force-free models with various boundary shapes (see their Fig.~10). For an elliptical boundary, this relationship is summarized as 
$\rBx(\ysb ,b,n=1)$. This applies in particular to the Lunquist field ($b=1$) and it is simply summarized as $\rBxL (\ysb) \equiv \rBx(\ysb ,b=1,n=1)$.
  
   Next, a similar $\Bxav\!\! / \!\!\!\Bav$ is expected when $\BL$ is fitted to $\vec{B}$ (since $\BL$ approaches the best possible $\vec{B}$).  Setting the equality $\rBx (\ysb ,b,n=1) = \rBxL (p)$ provides a relation $p(\ysb ,b)$.  For a fixed $\ysb$ value, the derivation of this relation implies: $\rmd p/\rmd b = (\rmd \rBx /\rmd b) / (\rmd \rBxL /\rmd p)$. Since $\rBx$ is a decreasing function of $b$ for a fixed $\ysb$ and $\rBxL$ is an increasing function of $p$ \citep{Demoulin09b}, this implies that $p$ is a decreasing function of $b$. 

Finally, with the magnitude of change of $\rBx$ with $b$ found in Fig.~10 of \citet{Demoulin09b}, the value of $b$ is expected to strongly affect the estimated $p$ value, so it is expected to strongly bias the MC probability distribution (such as shown in \fig{Pobs_OKCA}).  Such expectation is tested below by fitting with $\BL$ a variety of models with elongated cross section.      
     
\subsection{Models with elongated cross-section} 
\label{sec_elong_Models}

We explore the space of parameters mainly with analytical models as summarized in \fig{schema_cube}.    
The emphasis is set on the aspect ratio $b$ since it was found to be the most important parameter affecting $p$ (for a fixed $\ysb$).  We first analyze the model of \citet{Vandas03} who derived an analytical solution of a linear force-free field ($n=1$) contained inside an elliptical boundary 
(so generalizing $\BL$). 

   A numerical extension of the above model to cross sections with a bent (bean-like) shape was analyzed by \citet{Demoulin09b}.  They also consider the limit case of a rectangular cross section. It has a simple analytical expression for a linear force-free field ($n=1$): see their Eq.~(14), while their Eq.~(15) should rather be:   $\alpha_{\rm R}= \pi/2 \sqrt{1+b^2}/b$.
   
Finally, even if we have shown in \sect{cir} that $n$ has a small effect for circular cross section, we also consider the force-free field with $n=0.5$ and an elliptical cross section: 
  \BE
  (B_x,B_y,B_z) = \left(-\frac{y}{b \sqrt{1+b^2}},\frac{x~b}{\sqrt{1+b^2}},
                  \sqrt{1-x^2-\left( \frac{y}{b}\right)^2} \right)   \,. \label{eq_BJ}     
  \EE 
   
\begin{figure*}[t!]    
\centerline{\includegraphics[width=\textwidth, clip=]{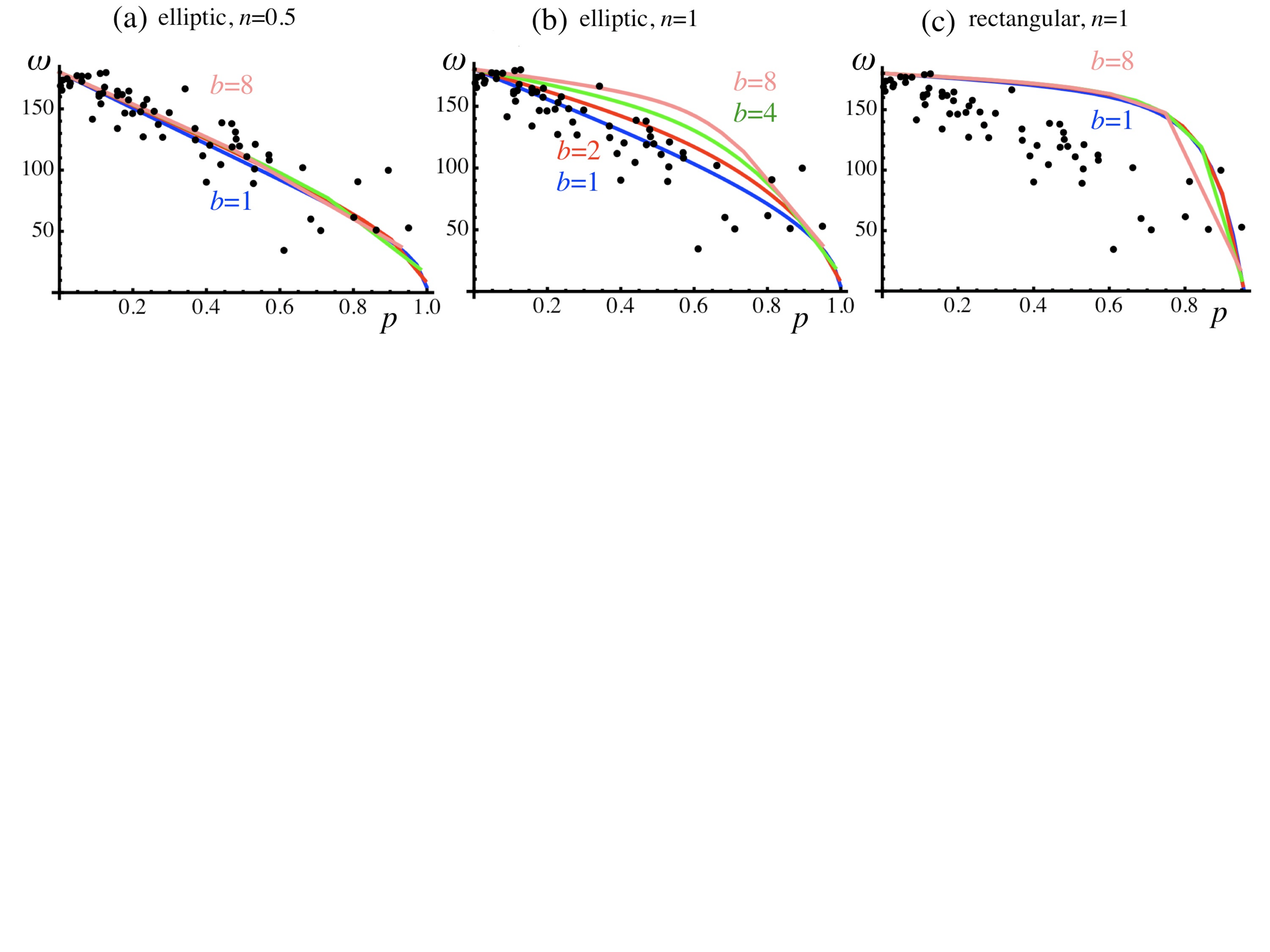}}
     \vspace{-0.48\textwidth}   
\caption{Dependence of the rotation angle, $\rotB$, of the magnetic field component orthogonal to the axis in function of the impact parameter, $p$, found by fitting the Lundquist field to two models with an elliptical boundary (a,b) and one with a rectangular boundary (c). The dots are the results obtained by \citet{Lepping10} for 65 MCs (see their Fig. 3B).   
}
\label{fig_rotEJR(p)}
\end{figure*}

\begin{figure*}[t!]    
\centerline{\includegraphics[width=\textwidth, clip=]{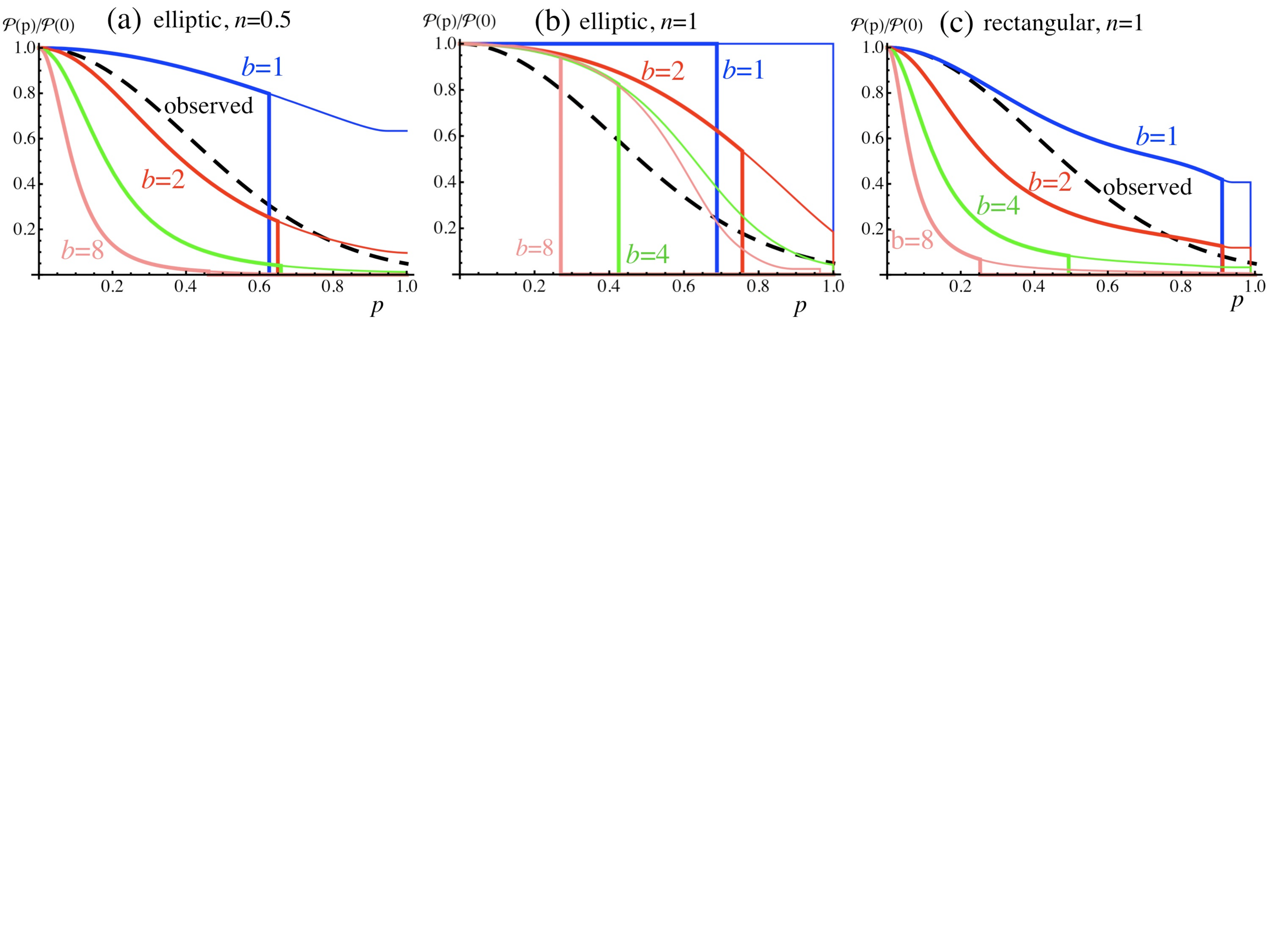}}
     \vspace{-0.5\textwidth}   
\caption{Probability distribution of the impact parameter, $\pp (p)$, deduced from various models and a uniform probability distribution of $\ysb$.  The thin curves are without selection effect and the thick curves are for $<B>/B_{\rm axis}>0.25$ and a magnetic field rotation angle greater than $90\degree$.
The dashed black curve is a Gaussian function fitted to observations (\fig{Pobs_OKCA}).
}
\label{fig_prob(p)}
\end{figure*}

\subsection{Effect of the aspect ratio} 
\label{sec_elong_fit}

  As the aspect ratio $b$ increases, more significant deviations between the synthetic observations of the modeled $\vec{B}$ and the fitted $\BL$ are present.  
For example, \fig{BfitsEJ} shows two extreme cases with $b=4$ and $\ysb =0.9$ (similar fits are obtained with lower $\ysb$ values). For both cases the field rotation angle, $\rotB$, is about $120\degree$, so large enough to be detected.  
However, using at typical value of $\Baxis=20$~nT at 1~AU, for $n=1$ such flux rope would not be detected, but for $n=0.5$ it would since the respective mean magnetic field strength along the simulated trajectory is $\leq 0.8$ and $\sim 9$~nT (compare to a typical SW field $\approx 5$~nT).
  
As expected in \sect{elong_aspect_ratio}, the aspect ratio $b$ has a strong effect on the estimated impact parameter $p$ (\fig{resEJ(p)}a,c).  This effect is much stronger than the effect of $n$ for circular flux ropes (\fig{resAn(p)}a).   The results obtained for $n=1$ and a rectangular boundary are similar (so not shown) to the results for $n=0.5$ and an elliptical boundary. 
All these results imply that $p$ is systematically biased to a lower value than the true impact parameter $\ysb$, and this effect strongly increases as $b$ is larger.
 
As with circular models (\sect{cir_info}), the quality of the fit of $\BL$ to the synthetic data, so a low value of $\dev / \!\!\!\Bav$, cannot be used to estimate the quality of the derived fitted parameters, in particular of $p$.
Indeed, even small values of $\dev / \!\!\!\Bav$ are present for large $p$ values (\fig{resEJ(p)}b,d) where $p$ is the most biased compared with $\ysb$ (\fig{resEJ(p)}a,c). 
Moreover, $\dev / \!\!\!\Bav$ has only a small dependence on $b$ for $n=0.5$ while $p$ has a strong dependence on $b$.  We conclude that the value of $\dev / \!\!\!\Bav$ is not a reliable way to qualify the best fitting model.
  
As the aspect ratio $b$ is increased, the flux rope is stretched in the $y$ direction, so one expects an increase of $\By$ to the expense of $\Bx$, so an increase of the field rotation angle $\rotB$.  In the models shown here, this enhanced rotation angle is mainly present for the elliptical case with $n=1$ (\fig{rotEJR(p)}b).  For the two other models the rotation angle is almost independent of $b$ (\fig{rotEJR(p)}a,c). 
  
With a larger $b$ value, the magnetic field can expand further away in the $y$ direction, implying lower field strength (see \eg\ Fig.~5 of \citealp{Demoulin09b}).  The fit of $\BL$ to this weaker field leads to a lower $\Bo$ (much lower than $\Baxis$).  Then, we found that $\Bo$ is a faster decreasing function of $p$ for larger $b$ values.  By contrast, $\Bav \!\!/\Bo$ is found to be almost independent of $b$ and more generally of the boundary shape. Then, the slightly higher value of $\Bav \!\!/\Bo$ for MCs than for a Lundquist field, as found by \citet{Lepping10} in their Fig.~5, is mainly related to the $\Bz (A)$ relation, and in particular to $n$, \eq{B_z(A)}, as found at the end of \sect{cir_info}.
  

\subsection{Expected observed distribution of impact parameter} 
\label{sec_elong_Expected}

The above bias on the estimated impact parameter, $p$, has important implications for the observed probability distribution (\eg\ \fig{Pobs_OKCA}).  More precisely, let us consider MC models with the same physical characteristics and observing bias (called $\cha$, which defines a set of five parameters, see Table~\ref{T-parameters}).  The simulated crossing is set at $\ysb$ with a distribution $\py (\ysb)$.   The models present in the interval $[\ysb, \ysb + \rmd (\ysb) ]$ are mapped to the interval $[p, p + \rmd p]$ with the Lundquist fit. The two probability distributions are related by: 
  \BE \label{eq_Pconserv}
   \pp (p,\cha ) ~\rmd p = \py (\ysb) ~\rmd (\ysb)  \,.   
  \EE

Moreover, some flux ropes could not be recognized as MCs because the crossing was too close to the flux-rope border.  We include two important selection effects: a too weak field strength and a too low rotation angle of the magnetic field.  
Other selection effects are associated to the presence of strong distortions, especially present when two MCs are interacting \citep[\eg ][]{Wang03,Lugaz05,Dasso09}.  Our models cannot take into account these relatively rare cases of MCs in interaction.  For isolated MCs, the distortions close to the boundary are expected to be the strongest \citep[weaker magnetic field and stronger effect of the surroundings, \eg ][]{Lepping07b}, therefore we select a relatively large minimum rotation angle, $\rotBmin = 90\degree$, while MCs are detected in observations with
a minimum rotation angle of $\approx 40\degree$ (\fig{rotEJR(p)}).  The flux rope can also be missed if its magnetic field strength is too weak. We select cases with $\Bav \! / \Baxis \geq \Bmin $ with typically $\Bmin =0.25$, since for a typical $\Baxis$ value of 20~nT this implies that $\Bav$ is comparable to the typical field magnitude in the solar wind at 1~AU.   Supposing a uniform distribution $\py (\ysb)$ (see \sect{Introduction}), and including the above selection effects in \eq{Pconserv}, the probability to detect a MC is:
  \BE \label{eq_P(p)}
   \pp (p,\cha ) = \left. \frac{\rmd (\ysb)}{\rmd p} 
      \right|_{<B>/\Baxis \,\geq \Bmin \;\&\; \rotB \geq \rotBmin}  \,.   
  \EE
  
Within the studied models (\fig{schema_cube}) the weakest bias of $p$ with increasing $b$ is obtained for the linear force-free model, $n=1$, with an elliptic cross section (\fig{resEJ(p)}c). It implies a moderate decrease of $\pp (p,\cha )$ with $p$ without selection effect ($\Bmin =\rotBmin =0$, see thin curves in \fig{prob(p)}b).  The selection effect with $\rotB$ is only present for large $p$ values and its effect decreases with $b$ (\fig{rotEJR(p)}b).  However, this model has also the weakest $\Bav$ for large $p$ values.  It implies a strong selection effect for $\Bmin =0.25$, increasing with $b$ (thick curves in \fig{prob(p)}b). 

Increasing the axial currents ($n=0.5$) or extending the cross-section to a rectangular shape both imply a stronger magnetic field for large $p$ so a weaker selection effect.   
The selection effect with the field rotation angle remains also limited to large $p$ values (\fig{rotEJR(p)}a,c).   
Moreover,  as $b$ is increased, the strong decrease of $\rmd (\ysb)/\rmd p$ with $p$ further decreases the selection effects (compare thin and thick curves in \fig{prob(p)}a,c).   
{\bf We conclude that the relation $p(\ysb )$ has generically a major effect on the observed MC distribution drawn in function of $p$.}

Could we interpret the observed distribution $\pobsp (p)$ (\fig{Pobs_OKCA}) as due to oblate cross-sections?   For the elliptic case with $n=1$, none of the $\pp (p,\cha )$ distributions, with a fixed $b$ and selection criteria, is close to the observed distribution (\fig{prob(p)}b). 
However a mixture of such distributions well could be, and this will be analyzed in \sect{aspect}.  
For the elliptic case with $n=0.5$, $\pp (p,\cha )$ is very close (\ie\ within the error bars) to the observed distribution $\pobsp (p)$ for $b\approx 2$ (\fig{prob(p)}a), while for the rectangular boundary with $n=1$, $\pp (p,\cha )$ is also close to the observed distribution for $b\approx 1.5$ (\fig{prob(p)}c).

\begin{figure}[t!]    
\centerline{\includegraphics[width=0.48\textwidth, clip=]{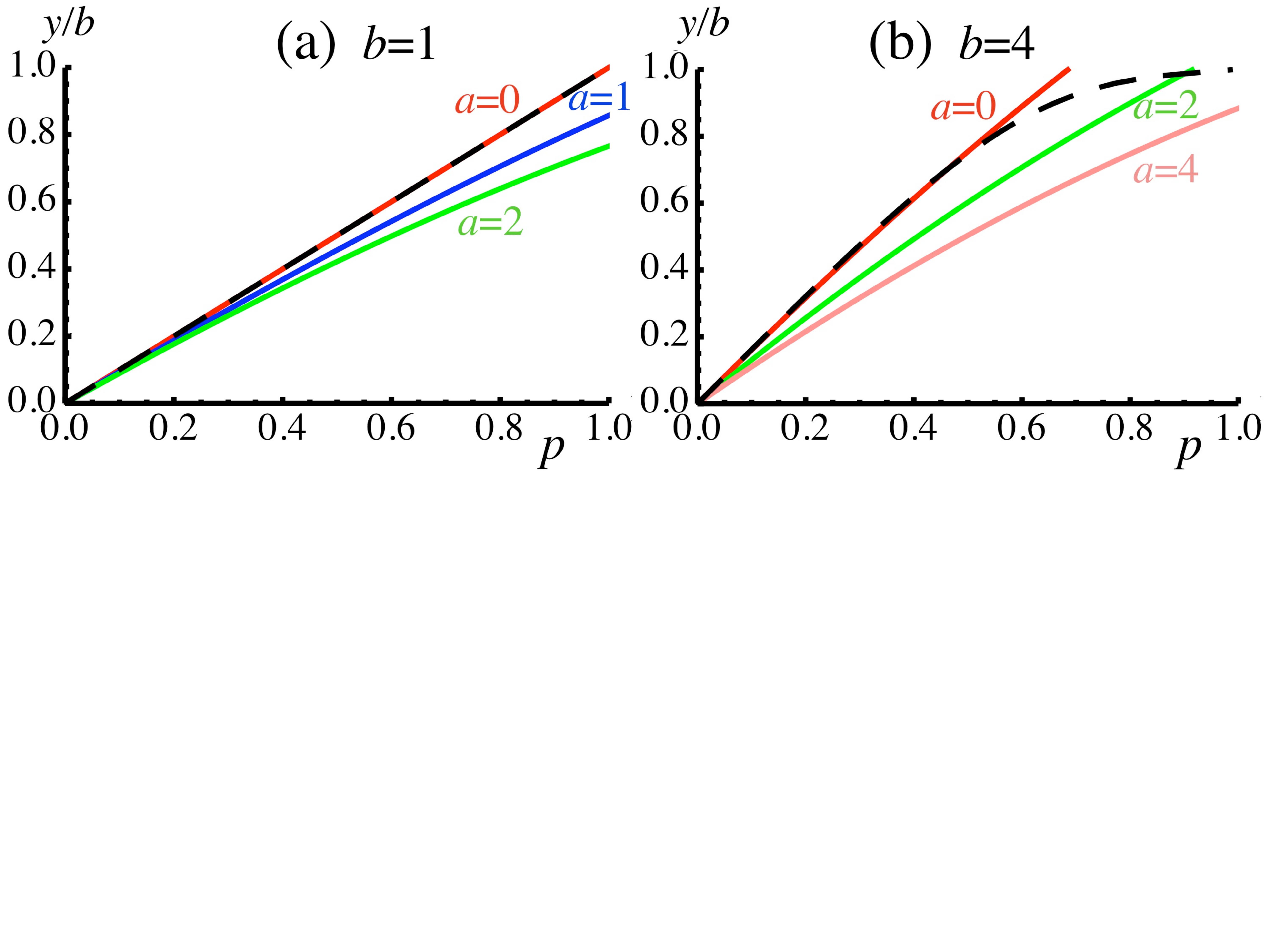}}
     \vspace{-0.18\textwidth}   
\caption{Approximate dependence of the true impact parameter, $\ysb$, in function of the estimated impact parameter, $p$, for bent cross-sections derived from \citet{Demoulin09b} results (derived from $\rBx$, see text in \sect{elong_bent}). 
The bending increases with the dimension-less parameter $a$.  Linear force-free models ($n=1$, \eq{B_z(A)}) are shown for two aspect ratio $b$.   The black dashed line is the relation 
found by fitting the Lundquist field to the elliptic ($a=0$) model with $n=1$. }
 \label{fig_ysb(p)}
\end{figure}

\begin{figure*}[t!]    
\centerline{\includegraphics[width=0.95\textwidth, clip=]{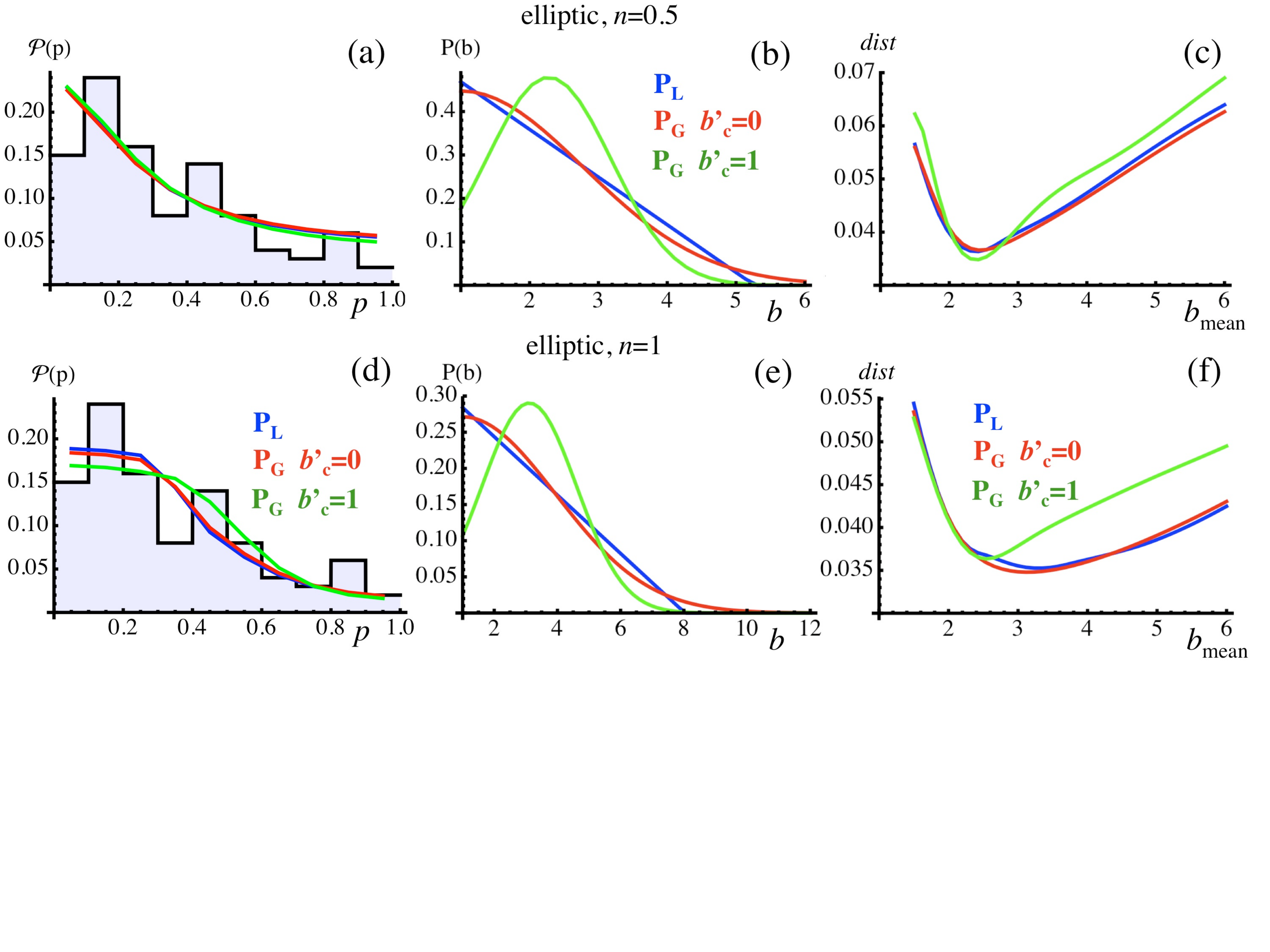}}
     \vspace{-0.2\textwidth}   
\caption{Probability distributions $\pp (p)$, from \eq{P(p)int}, and $\pb (b)$, from \eqs{pbL}{pbG}, for the minimum of $\dist(\eta,\bmean)$ as defined by \eq{dist}. $\pobsp (p)$ of \fig{Pobs_OKCA} is added in (a,d).
Two force-free elliptical models are shown: {\bf (a-c)} $n=0.5$ (constant current density), {\bf (d-f)} $n=1$ (linear force-free) for $\Bmin=0.25$ and $\rotBmin =0$.  The three $\pb (b)$ functions, shown with three colors, imply similar results.   
}
\label{fig_prob_dist}
\end{figure*}

\subsection{Effect of a bent cross-section} 
\label{sec_elong_bent}

 In some MHD simulations, the flux rope is strongly compressed in the propagation direction, such that its front region becomes relatively flat \citep[\eg ][]{Vandas02b}. The cross section can even develop a bending of the lateral sides towards the front direction when it central (resp. lateral) parts move in a slow (resp. fast) solar wind (\eg , \citealt{Riley03}; \citealt{Manchester04b}).  Then, \citet{Demoulin09b} have investigated the effects of bending the flux-rope boundary to a bean-like shape.
This bending is parametrized by the dimension-less parameter called $a$. Examples of computed fields with various $a$ values are shown in their Figs.~3-5.   Typically $|a|$ needs to be larger as $b$ increases to get a comparable bending.  

Does bending of the flux-rope cross section modify the probability distribution of flux rope detection, $\pp (p,\cha )$, versus the estimated parameter $p$?  The effect of $a$ value can be approximately derived from the relation $\rBx\!\! =\; \Bxav \!\! / \!\!\!\Bav $ in function of $\ysb, a$ and $b$ as summarized by the analytical expression of Eq.~(31) of \citet{Demoulin09b}.  As in \sect{elong_aspect_ratio}, a similar $\Bxav\!\! / \!\!\!\Bav$ is expected for $\vec{B}$ and its fitted $\BL$ field.  Setting the equality $\rBxL (p)=\rBx (\ysb ,a,b)$ provides an estimation of $p$, named $p_{\rm e}(a,b,\ysb )$, which is shown in \fig{ysb(p)} for a few $a$ and $b$ values.  

We also compare the estimation $p_{\rm e}$  to the result of fitting $\BL$ to $\vec{B}$ for $a=0$, so an elliptical boundary (dashed black line).  For $b=1$, both curves are simply $p=p_{\rm e}=\ysb$  (\fig{ysb(p)}a), while for $b>1$ there is a good agreement up to large $y/b$ values (\fig{ysb(p)}b). Such result could be extended to $a>0$ by applying the Lundquist fit to the bent models
developed by \citet{Demoulin09b}.  Then, the analytical expression $p_{\rm e}(a,b,\ysb )$ provides an estimation of $p$ for a broad range of $\{a,b,\ysb \}$ values. 
This result has a practical application: it provides a good initial guess of $p$ (from observed $\rBx$) for the non-linear fit of $\BL$ to $\vec{B}$ (so both avoiding to start in a wrong local well of \dev , \eq{dev}, and speeding up the computations).  
   
Figure~\ref{fig_ysb(p)} shows that bending of the flux-rope cross section, so increasing $|a|$, increases $p$ for a given $\ysb$.  This is the opposite effect to increasing $b$ (\fig{resEJ(p)}a,c).  From \eq{P(p)}, this implies a bias increasing the probability of flux ropes for large $p$, so the opposite to observations (\fig{Pobs_OKCA}).   It is also worth noting that $|a|=2$ is already a very bent cross section \citep[see Figs. 4,5 of][]{Demoulin09b} that we expect to be rarely present in observed MCs.  
We conclude that the effect of bending the cross-section is expected to introduce only a weak bias to the estimated $p$ value.

\section{Distribution of the cross-section aspect ratio} 
\label{sec_aspect}

In the previous section we found that the probability distribution $\pp (p,\cha )$, was most sensitive to the aspect ratio $b$.
In this section we use this property to constrain the probability distribution, $\pb (b)$, of the aspect ratio $b$ for the MCs observed at 1 AU.   We end by exploring how $\pb (b)$ depends on the MCs properties. 

\subsection{Method} 
\label{sec_aspect_Method}

In the following, we consider that the aspect ratio, $b$, is distributed according to the probability function $\pb (b)$, while the other parameters in $\cha$ remain the same.  
Because of the small effect of $a$, see \sect{elong_bent}, we set $a=0$.  The expected probability of the impact parameter $\pp (p)$ is the superposition of the contribution of each $b$ values according to:
  \BE \label{eq_P(p)int}
   \pp (p) =   \int_{\bmin}^{\infty} \pb (b) \; \pp (p,\cha ) \;\rmd b \,,   
  \EE
where $\cha=\{a,b,n,\Bmin,\rotBmin \}$ and $\int_{b_{\rm min}}^{\infty} \pb (b) \;\rmd b =1$, while 
$\int_0^1 \pp (p,\cha ) \;\rmd p \leq 1$ since cases are missed with the selection on $\Bmin$ and $\rotBmin$. 
 
Since $\pb (b)$ is contributing through an integral to the distribution $\pp (p)$ in \eq{P(p)int}, and that $\pobsp (p)$ (shown in \fig{Pobs_OKCA}) has important uncertainties due to the limited number of observed MCs, we can only derive a global behavior of $\pb (b)$.  For that, 
we limit the freedom of $\pb (b)$ by selecting functions which depend on few parameters ($\bmean , b'_c$), and we minimize the distance, $\dist$, between $\pp (p)$ and $\pobsp (p)$:
  \BE \label{eq_dist}
   \dist = \sqrt{ \int_0^1 \left( 
       \pobsp (p) - \eta \int_{1}^{\infty} \pb (b) \; \pp (p,\cha ) \;\rmd b
       \right)^2 \; \rmd p}  \;. 
  \EE
We introduce the parameter $\eta$ in front of $\pp (p)$ since $\pobsp (p)$ is normalized with all the observed MCs ($\int_0^1 \pobsp (p) \;\rmd p = 1$) while each $\pp (p,\cha )$ is normalized to all the cases. Since we cannot also normalize $\pobsp (p)$ to all cases, we let $\eta $ as a free parameter.  It is expected to be around $1$ since the selection biases are expected to be small (\sect{elong}).  

The generic cross-section shape of MCs is mostly unknown since the only shape determinations were done with a Grad-Shafranov reconstruction technique or by fitting the elliptical linear force free model on a few MCs (see \sect{Introduction}).   The cross section has the tendency to be round ($b\approx 1$) because of the magnetic tension and the typically low plasma $\beta$ found in MCs. However, the large pressure of the MC sheath tends to elongate the cross-section orthogonally to the MC mean velocity, so $b>1$.   Then, we set a minimum value for $b$ as $\bmin =1$.
Indeed, the MCs with $b<1$ cannot be too numerous, otherwise more MCs with large $p$ would be observed (because for $b<1$ the bias of $p(\ysb)$ is reverse than for $b>1$). 

We first select a simple linear function for $\pb (b)$:
  \BA \label{eq_pbL}
   \pbL (b,\bmean) &=& \frac{2}{(\bmax -1)^{2}} \;(\bmax -b) \quad {\rm if} \; 1 \leq b \leq \bmax \\
                   &=& 0 \hspace{0.18 \textwidth} {\rm otherwise,} \nonumber  
  \EA
where the coefficient in front of $(\bmax -b)$ is computed from the normalisation 
$\int_{1}^{\bmax}\pbL (b,\bmean) \;\rmd b =1$, and
  \BE \label{eq_bmean}
   \bmean = \int_{1}^{\bmax} b\; \pbL (b,\bmean) \;\rmd b = \frac{\bmax+2}{3}  \;. 
  \EE

As a second possibility for $\pb (b)$ we select a Gaussian distribution, limited to $b\geq 1$:
  \BA 
   \pbG (b,\bmean ,b'_c) &=& f \; \exp \left(-\frac{(b-b_c)^2}{2 \sigma^2} \right) 
                                                         \;, \label{eq_pbG}\\
{\rm with}\qquad f &=& \sqrt{\frac{2}{\pi}}\; \frac{1}{(1+\erf (b'_c)) \;\sigma} \;, \nonumber \\
              b'_c &=& (b_c-1) / (\sqrt{2} \;\sigma )  \;, \nonumber \\
            \bmean &=& \sqrt{2}\;\sigma 
                       \left(b'_c + \frac{\exp (-{b'}_c^2)}{\sqrt{\pi}\; [1+\erf (b'_c)]} \right) 
                       +1     \;, \nonumber 
  \EA
where $\erf$ is the error function. The coefficient $f$ was computed from the normalisation $\int_{1}^{\infty} \pbG (b,\bmean ,b'_c) \;\rmd b =1$.
The parameter $\bmean$ is the mean value of $\pbG$, restricted to $b\geq 1$. 
The freedom of $\pbG$ is expressed in function of $\{\bmean ,b'_c\}$ rather than with the usual parameters of a Gaussian distribution $\{b_c,\sigma\}$ (\eq{pbG}) in order to easily compare our results with the linear distribution $\pbL$ (\eq{pbL}). Moreover, as shown below, $\bmean$ value is the most stable result deduced from minimizing the function $\dist$ (\eq{dist}). Therefore, we set the parameter $\bmean$ in both distributions.  For a given $\bmean$, the normalized parameter $b'_c$ determines the location of the maximum of $\pbG$ and the spread of the distribution, as follows. 
The probability at $b=1$ divided by the maximal one, at $b=b_c$, is simply $\exp (-{b'}_c^2)$ so $b'_c$ describes how much the function $\pbG$ is peaked ($b_c=b'_c \sigma \sqrt{2} + 1$, then for $b'_c=0$, its maximum is at $b=1$, while it is more peaked toward $b>1$ as $b'_c$ increases).
  
\subsection{Probability distribution of aspect ratio: $\pb (b)$} 
\label{sec_aspect_Prob}

  In this section, for a given $\pobsp (p)$ (\fig{Pobs_OKCA}), the function $\dist (\eta,\bmean)$, defined by \eq{dist}, is minimized.  We provide typical results for $\pb (b)$.

The function $\dist (\eta,\bmean)$ has a well defined global minimum in all explored cases, see \eg  \fig{prob_dist}c,f where cuts through the minimum is shown in function of $\bmean$. For $n=0.5$ the minima are nearly at the same location ($\eta \approx 0.91 \pm0.01$, $\bmean \approx 2.29 \pm 0.01$ for the three $\pb (b)$ functions shown, while for $n=1$ $\eta$ is larger ($\approx 1.26 \pm 0.06$) and $\bmean$ is more broadly distributed (from $\approx 2.2$ to $3.4$).  For each $n$ value, the derived $\pp (p)$ are all very close and fit globaly well the observations (\fig{prob_dist}a,d), with a comparable minimum of $\dist$ ($\approx 0.036$ for $n=0.5$ and $\approx 0.035$ for $n=1$).  There are still some differrences: for the case $n=1$, $\pp (p)$ is slightly lower than $\pobsp (p)$ for both small and large $p$ values ($p<0.3$ and $p>0.7$), while it is the opposite for the case $n=0.5$ (\fig{prob_dist}a,d).   It is an indication that $n$ is typically in between these values in MCs, in agreement with the result found for $\Bav \!\!/\Bo(p)$ at the end of \sect{cir_info}.

\begin{figure}[t!]    
\centerline{\includegraphics[width=0.5\textwidth, clip=]{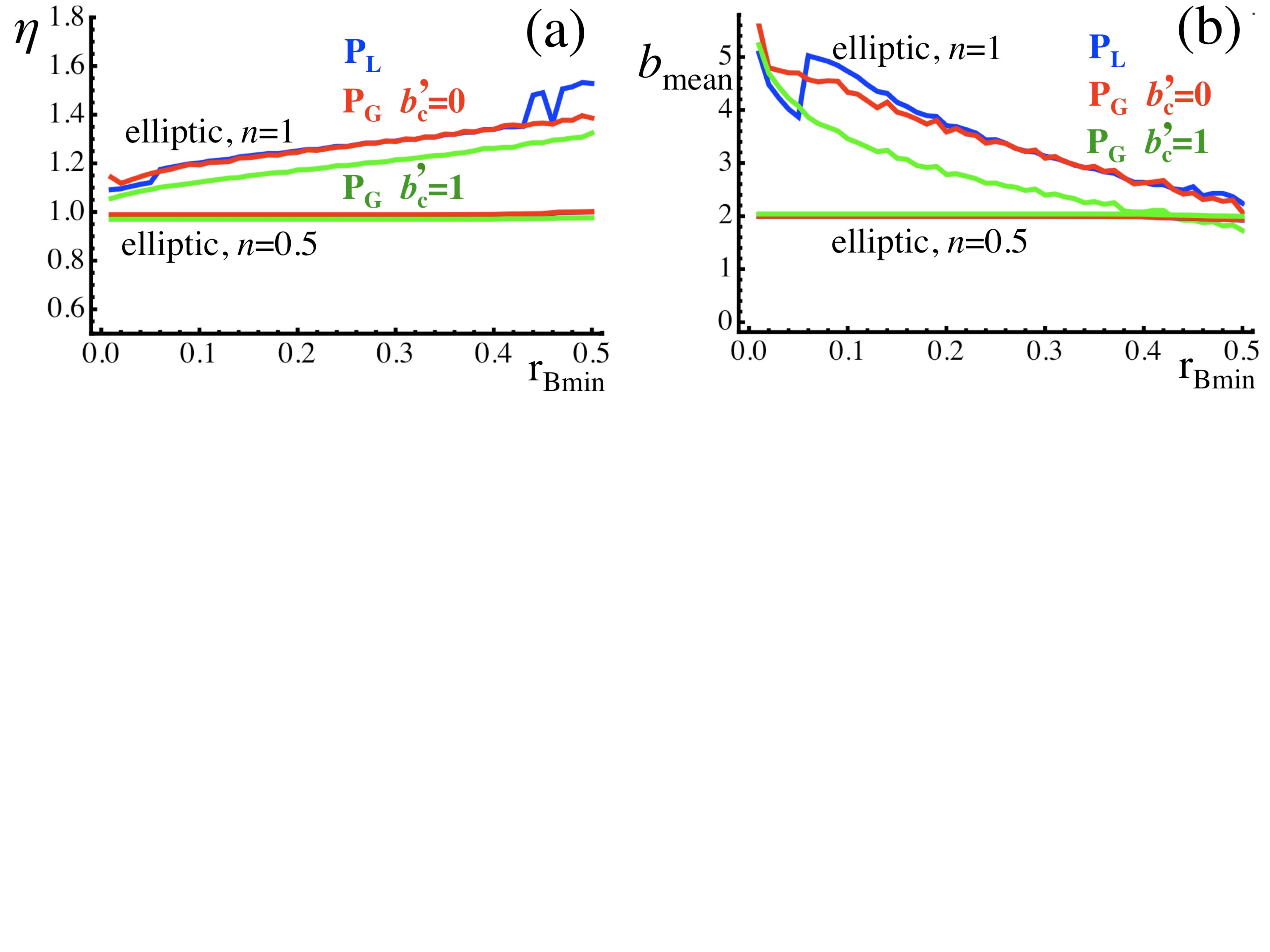}}
     \vspace{-0.22\textwidth}   
\caption{Effect of $\Bmin$ (the selection is defined by $<B>/B_{\rm axis}>\Bmin$ along the simulated crossing).  The coefficient $\eta,\bmean$ are found by minimizing $\dist$ (\eq{dist}) for two force-free fields (lower curves: $n=0.5$, upper curves: $n=1$).  Three probability distributions of $\pb (b)$ are shown with different colors (for $n=0.5$, the three curves are almost identical). 
}
\label{fig_c1Bmean}
\end{figure}

\begin{figure*}[t!]    
\centerline{\includegraphics[width=\textwidth, clip=]{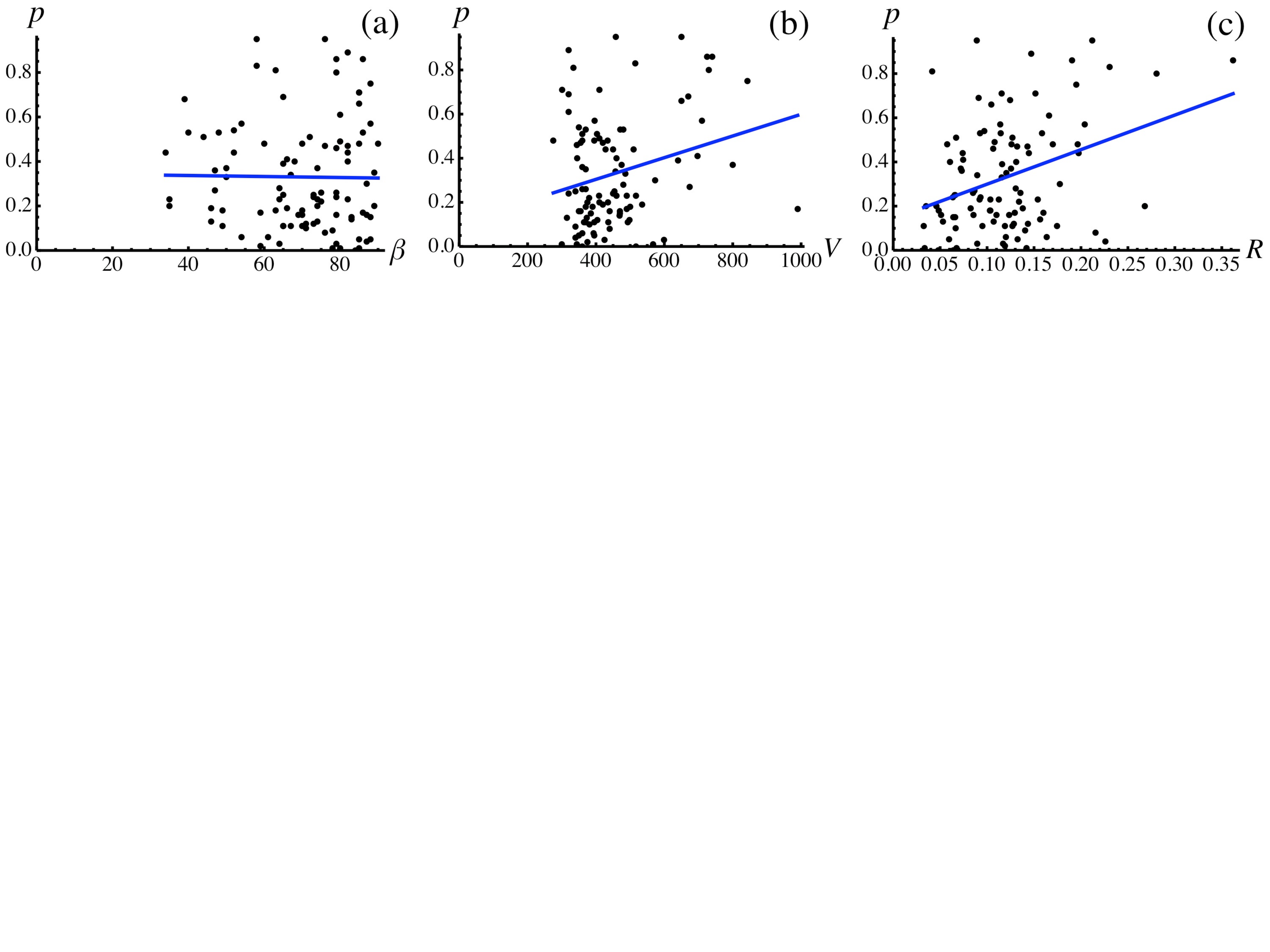}}
     \vspace{-0.53 \textwidth}   
\caption{Correlations of the impact parameter $p$ with: (a) the angle between the MC axis and the radial solar direction, (b) the mean MC velocity (in km/s), and (c) the flux-rope radius (in AU) found with the Lundquist fit.   
The straight line is a linear fit to the data points (MCs). }
\label{fig_correl}
\end{figure*}

\subsection{Sensivity of $\pb (b)$} 
\label{sec_aspect_Sensivity}

   We compare below the results for $\pbL$ and $\pbG$ 
varying both the models ($n$, cross-section shape) and the selection effects ($\Bmin$ and $\rotBmin$). 

  The results above are derived by fitting the theoretical results to $\pobsp (p)$ which has statistical fluctuations with the relatively low number (100) of MCs available.   
Then, we also derive the results from the Gaussian and linear fits (\fig{Pobs_OKCA}).
The larger change is present for the case $n=1$, and we find that $\bmean$ is inside the range $[2.2,3.4]$ for the $\pb (b)$ distributions shown in \fig{prob_dist}e.  The range found for $\bmean$ is changed to $[2.7,3.0]$ when the Gaussian fit is used, and to $[2.3,2.5]$ for the linear fit. 
For $n=0.5$, the changes are more limited: $\bmean \approx 2.29$ with $\pobsp (p)$, changing to $\approx 2.33$ for the Gaussian fit and $\approx 2.11$ for the linear fit. We conclude that the results are weakly dependent on the details of the function $\pobsp (p)$.

The selection on rotation angle, $\rotBmin$, has a low effect on the minimum of $\dist (\eta,\bmean)$ for $\rotBmin \leq 90\degree$.  The main effect of increasing $\rotBmin$ is to force $\pp (p)$ to zero for large $p$ values (\fig{prob(p)}).  This effect remains in the integration on $b$ in \eq{P(p)int}.  For example, with $\rotBmin = 90\degree$, $\pp (p)=0$ for $p>0.75$ for both $n=0.5$ and $1$, in contradiction with $\pobsp (p)$ (\fig{prob_dist}).  However, when $\rotBmin$ is decreased to $\approx 45\degree$, there is only a slight decrease of $\pp (p)$ for $p>0.9$, then $\rotBmin$ around $45\degree$ is compatible with $\pobsp (p)$ in agreement with the minimum rotation angle detected in MCs (\fig{rotEJR(p)}).

   We next explore the sensitivity of the results with $\Bmin$ selection.
The elliptical linear force-free field ($n=1$) is the most affected by changes of $\Bmin$ threshold (\fig{c1Bmean}).  This is indeed expected from the results of \sect{elong_Expected}, and in particular from what is shown in \fig{prob(p)}b.  
 As $\Bmin$ increases, so does the selection effect for large $p$ values, and lower $b$ values are needed to fit the observations and a larger $\eta$ is needed to compensate the selection effect (\fig{c1Bmean}).    At the opposite, the case $n=0.5$ is almost independent of $\Bmin$ since the selection affects only the low probability tail of $\pp (p,\cha )$, see \fig{prob(p)}a.   Similar results are obtained for $n=1$ and a rectangular shape, with only a shift of $\bmean$ to $\approx 1.56 \pm 0.01$ and $\eta$ increasing a bit to $1.07$, as expected from \fig{prob(p)}c.       

\textbf{We conclude that the observed probability $\pobsp (p)$ is mostly affected by the oblateness, $b$, of the flux-rope cross section.}
    
\subsection{Main constrain on $\pb (b)$} 
\label{sec_aspect_constrain}

  The results above are also relatively independent on the function $\pb (b)$ selected within the explored set.  
In all cases close results are obtained from a linear and Gaussian distribution having a maximum at $b=1$ (\eg \fig{prob_dist}).   
Moreover, similar results are found for a Gaussian distribution more peaked around its maximum, especially for the elliptical $n=0.5$ and the rectangular $n=1$ cases, 
\ie\ changing $b'_c$ has nearly no effect on $\eta$ and $\bmean$ values minimizing $\dist (\eta,\bmean )$.  This is illustrated by the cases $b'_c=0$ and $1$ in \figs{prob_dist}{c1Bmean}.  This is also true for much larger $b'_c$, so more peaked Gaussian function (indeed also in the limit $b'_c \rightarrow \infty$, so when $\pbG$ select only $b=\bmean$).   This property is linked to the behavior of the functions $\pp (p,\cha )$: the ones for $b \approx \bmean$ approximately fit the observations while the ones for larger $b$ are too peaked to low $p$ values and the opposite for lower $b$ values (\fig{prob(p)}a,c).  Then, for a distribution of $b$ values, the best fit is always found around the same $\bmean$ value, and the behavior of $\pp (p,\cha )$ for lower $b$ values tend to compensate those for higher $b$ values.   
  
  The above results can be modeled with the following analytical functions:
  \BE \label{eq_Panal}
   \pp (p,{\rm anal.} )  =  \frac{2 \sqrt{q~b/\pi}}{{\rm erf}(\sqrt{q~b})} \exp (-q p^2 b)  \,,   
  \EE
which approximate the behavior of $\pp (p,\cha )$ for the $n=0.5$ elliptical case with $q\approx 1.4$, and for the $n=1$ rectangular case with $q\approx 2$.  The $n=1$ elliptical case has $\pp (p,\cha )$ functions the most different from $\pp (p,{\rm anal.} )$, while still with some global similarities in their dependences with $p$ and $b$.  Then, and even in this case the results are weakly dependent on $b'_c$ (\fig{c1Bmean}). We conclude that the observations, summarized with $\pobsp (p)$, mainly determine the mean value of $b$, independently of the shape of $\pb (b)$.

\section{Application to subsets of MCs} 
\label{sec_subsets}
   
\subsection{Correlation between MC parameters} 
\label{sec_aspect_Correlation}

 In this section we explore the correlations between $p$ and the other global parameters measured in the set of 100 MCs observed at 1~AU.   In particular we find unexpected correlations. 
  
First, we examine the cone angle $\beta$ which is the angle between the MC axis 
to the solar radial direction.  The number of detected MCs decreases with a lower $\beta$ angle (\fig{correl}a).  Still, we find no correlation between $\beta$ and $p$, showing that the crossing cases away from the MC nose (low $\beta$ values) have no special biased impact parameter when the few cases corresponding to a leg crossing are filtered out ($\beta>30\degree$). 
This justifies the use of models with locally straight axis \citep[\eg ][ and references therein]{Owens12}.   

As reported by \citet{Lepping10}, we also find no significant correlation between $p$ and $\Bo$ (the deduced axial field strength).   We agree with their interpretation that $\Baxis$ of MCs is expected to be spread in a large range (about a factor 10), so that the dispersion of $\Baxis$ is likely to mask any weak dependence $\Bo(p)$.  Indeed, we find such dependence in the models.  
The dependance is weak for circular models (\fig{resAn(p)}) and moderate for models with elongated cross sections.
For example, with $b=2$, $\Bo$ monotonously decreases from $1$ to $\approx 0.4$ for an elliptical linear-force free field, while this decrease is much weaker, only down to $\approx 0.8$, for an elliptical model with uniform current (not shown).  Since we found two indicators in favor of finding typical MCs in between those models (ends of \sects{cir_selection}{elong_Models}), $\Bo(p)$ is expected to have a relatively weak dependence (from $1$ to $\approx 0.6$) which can be easily masked by the large dispersion of $\Baxis$ in MCs.

\begin{figure}[t!]    
\centerline{\includegraphics[width=0.4\textwidth, clip=]{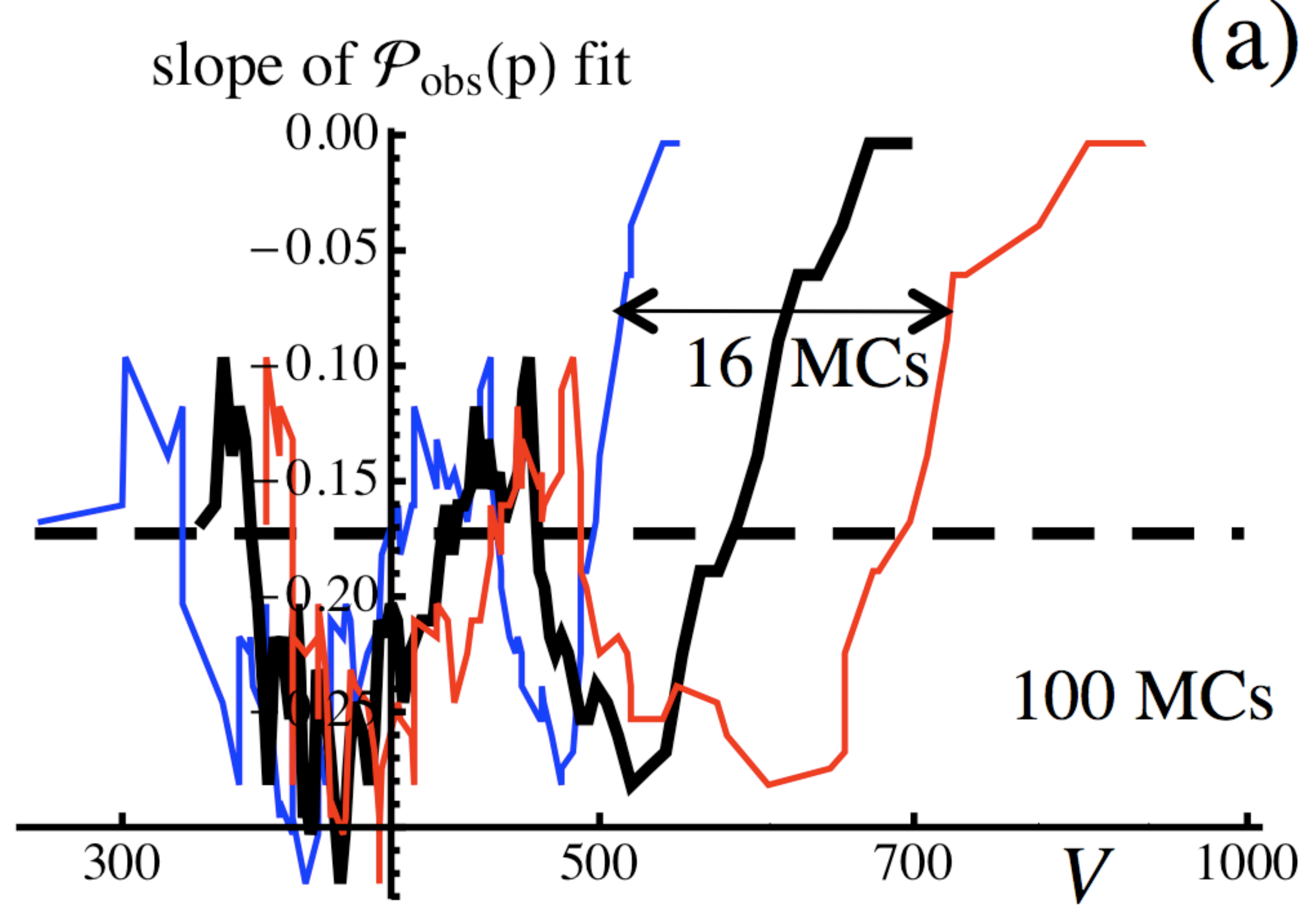}}
\centerline{\includegraphics[width=0.4\textwidth, clip=]{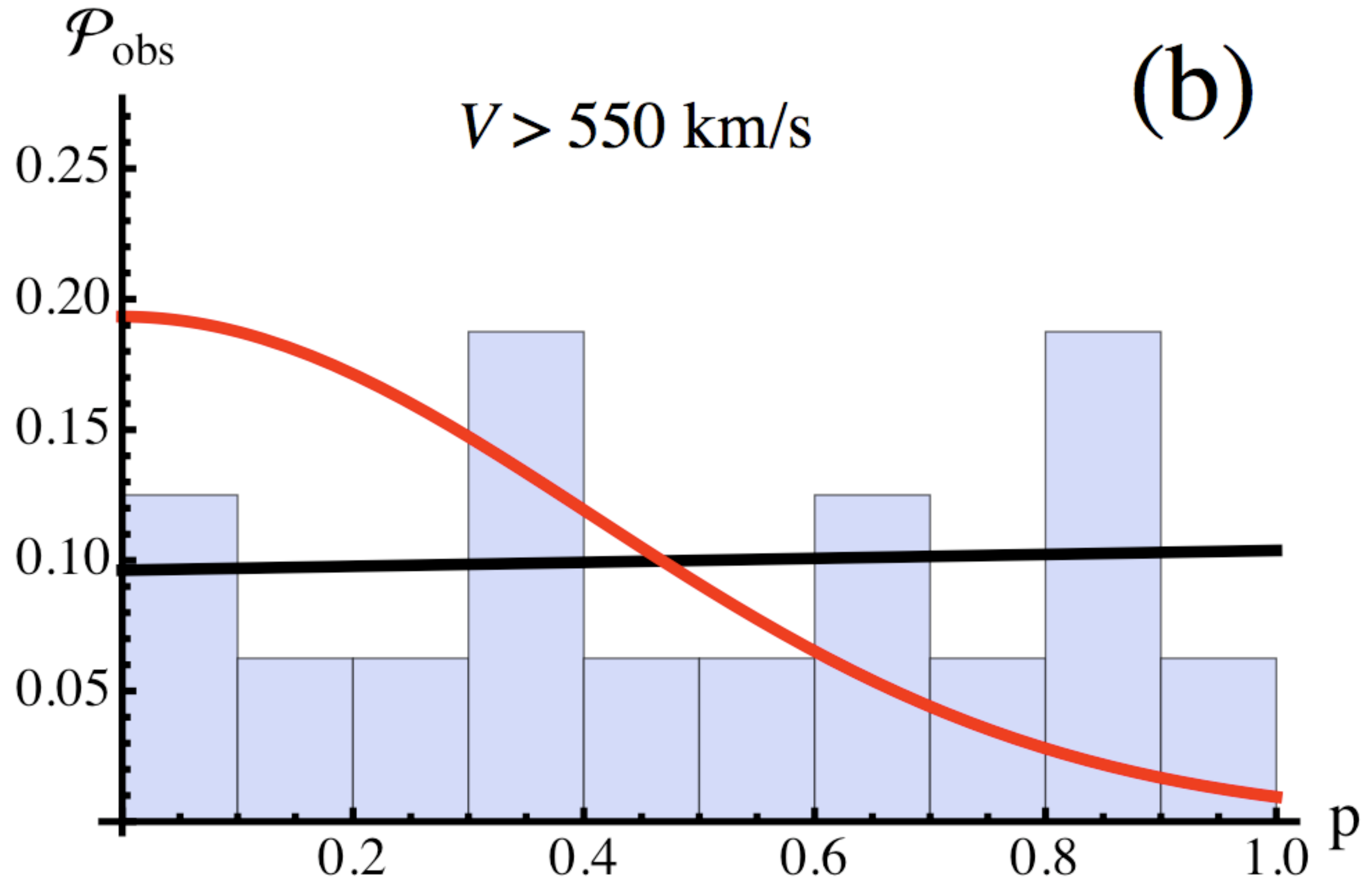}}
\centerline{\includegraphics[width=0.4\textwidth, clip=]{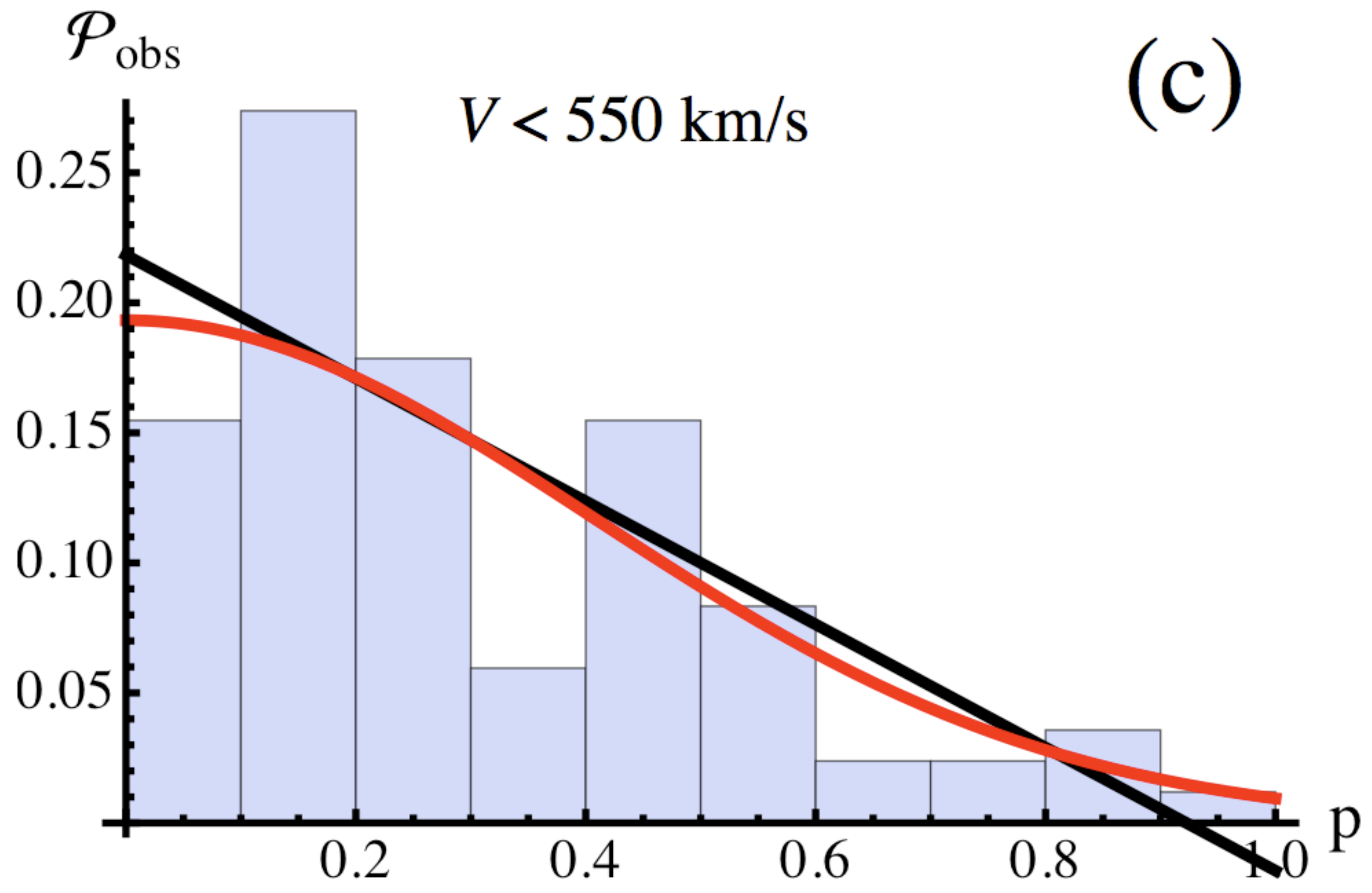}}
\caption{ Properties of impact parameter distributions $\pobsp (p)$ for different MC groups.
  {\bf (a)} Slope of the linear fit of $\pobsp (p)$ when MCs are first ordered with a growing mean velocity $V$, then binned in groups of 16 cases (running grouping with increasing $V$). The three curves represent this slope, with the black line corresponding to the mean V of each group, and the blue (resp. red) line corresponding to the minimum (resp. maximum) value of each group.  The horizontal dashed line is the slope for all MCs (slope of the black line in \fig{Pobs_OKCA}).  The horizontal axis is scaled with the logarithm of $V$.
  {\bf (b,c)} Probability distribution, $\pobsp (p)$, as in \fig{Pobs_OKCA}, with MCs separated in two groups according to their mean velocity $V$.
}
\label{fig_sel_V_R}
\end{figure}

Other global parameters are not or only weakly correlated with $p$ except two: $V$ (mean velocity of the MC along the spacecraft trajectory) and $R$ (flux rope radius deduced from the Lundquist field).  The Pearson's correlation coefficient is $0.26$ and $0.35$ for $V$ and $R$ respectively, and a linear fit also clearly show the trends (\fig{sel_V_R}b,c).  
The correlation $V(p)$ is the most surprising since $V$ is measured directly from the data and it is a robust quantity (weakly dependent on the selected MC boundaries).   
Such result could not be interpreted as a real velocity shear between the MC core and its surrounding since by its magnitude this effect would shear apart the flux rope before its arrival to 1~AU 
(and consequences of this particular behavior around 1~AU is both not observed and not plausible).
The strong correlation $R(p)$ is also surprising.  Still, we emphasis the study of $V$ because $R$ could be affected by the amount of reconnection achieved between the MC and the overtaken magnetic field as deduced by the presence of a back region in MCs \citep{Dasso06,Dasso07,Ruffenach12}.

\subsection{Sets of MCs with different aspect ratios} 
\label{sec_aspect_Sets}

We investigate the above puzzling result by analyzing probability distributions, as in \fig{Pobs_OKCA}, but for MCs with a restricted interval of velocity.   
Due to the relatively low number of MCs, we are limited to a relatively coarse 
sampling in $V$. 

The probability distributions are fitted by a straight line (as the black line in \fig{Pobs_OKCA}) in order to decrease the statistical fluctuations inside the $p$ bins and to summarize the distribution information to the slope.  For an histogram of $N$ MCs and distributed according to a linear function of $p$, the constrain that the sum of the probabilities is unity implies a relation between the slope of this distribution and the mean value of $p$, noted $<p>$, as:
  \BE  \label{eq_slope}
  {\rm slope} = 12 ~\Delta p ~(<p>-1/2) / (1-N^{-2}) \,, 
  \EE
with $\Delta p$ being the bin size. For $N$ slightly large (say $N\geq 10$), \eq{slope}
shows that the slope is almost independent of $N$ and simply related to $<p>$.
It implies that the slope is a relatively robust quantity, even for a low number $N$ of MCs used to build the distribution. The expected statistical fluctuations on $<p>$ are of the order of $<p>/\sqrt{N}$, which translate to fluctuations of the slope $\approx ({\rm slope} + 6 \Delta p)/\sqrt{N} \approx 0.1$ for a slope $\approx -0.2$ (\fig{Pobs_OKCA}), $\Delta p =0.1$ and $N=16$.  

Next, we ordered the MC data according to growing values of $V$, and computed the evolution of the slope for $N$ MCs progressively shifting to higher $V$ values. With $N=16$, fluctuations of the slope are $\leq 0.1$, as expected.  There is a sudden change for $V$ above $\approx 550$ km/s (\fig{sel_V_R}a).  A similar result is obtained for larger $N$ values, with less fluctuations, but with a reduced dynamic (in both axis directions).  Indeed, separating the MCs in two groups shows two different distribution functions (\fig{sel_V_R}b,c).  Similar results are found when the above ordering with $V$ is replaced by one with $R$. 
  
With the results of \sects{aspect_Prob}{aspect_constrain}, we interpret this result as the presence of two main groups of MCs.  The slower ones, $V<550$ km/s, which are also the most numerous (84 MCs), have an oblate cross section with a mean aspect ratio between 2 and 3 depending the model used, similar to the full set of MCs.   However, the faster MCs at 1~AU have a nearly flat distribution, so they are mostly round whatever model is selected (within the explored ones).  It would be worth checking this conclusion with more MCs since this group is limited to 16 MCs.   These  MCs are also typically larger and with stronger magnetic field since $V$ has a correlation coefficient of 0.32 with $R$ and 0.68 with $\Bo$ (for the full set of 100 MCs). 
Indeed, a variation of the slope of $\pobsp (p)$ with MCs ordered with $R$ was found to be similar than with $V$ (\fig{sel_V_R}a).  This is not the case with $\Bo$ since there is no significant correlation between $\Bo$ and $p$ (\sect{aspect_Correlation}). 

Why faster and larger MCs would have typically nearly round cross section?  At first thought, a faster MC would imply a larger snowplow effect, plausibly generating a larger sheath which can compress more the flux rope, inducing a flatter cross section.  
However, the velocity is measured at 1~AU and the above result could rather mean that those faster MCs were in average less decelerated than others, so that the distortion from the surrounding solar wind was less important than for other MCs.
Moreover, faster MCs spend less time from solar eruption to their arrival to the point where they 
are \insitu\  observed, and then distortion mechanisms are expected to be less effective to operate.
Another, and plausibly complementary answer is that the faster MCs have typically a stronger magnetic field, so that the magnetic tension is stronger and keep the cross section more round. 

\section{Conclusions} 
\label{sec_Conclusion}

The MCs observed at 1~AU are classically fitted with a Lundquist model \citep{Lepping90}. In the set of 120 MCs analyzed, only 11\% (13/120) of the MCs could not be satisfactory fitted (either the flux-rope handedness could not be determined, either the fit was not converging), while 6\% (7/120) of the MCs are crossed to far away from the nose to provide reliable fit results.  For the remaining 100 MCs the fit provides an estimation of the impact parameter ($p$).  
The observed probability distribution, $\pobsp (p)$, of these MCs is found to decrease strongly with $p$ \citep[][ and \fig{Pobs_OKCA}]{Lepping10}.
Compared to an expected almost uniform distribution, this could imply that about half of the MCs are not detected by \insitu\ observations.  
Is this decrease due to a strong selection effect, like on the magnetic field strength and/or the amount of field rotation angle? Or perhaps are the MCs observed only in about one third of ICMEs because more criteria are used to define ICMEs than MCs? Moreover, several of the less restrictive criteria used to identify ICMEs are expected to be independent of the impact parameter (such as temperature, composition and ionization level).  In order to answer these questions we explore the parameter space of flux rope models with force-free fields.  
We simulate spacecraft crossings and perform a least-square fit of the synthetic data with a Lundquist field, using the same procedure as for observations of real MCs.
The fit provides an estimated impact parameter $p$ that we compare to the true one 
known from the synthetic model.

For models with circular cross sections, we found that selection effects with magnetic field strength and field rotation angle are present only for large $p$ values, so they cannot explain the gradual decrease of $\pobsp (p)$.  This result is found for a broad variety of magnetic field profiles ranging from nearly uniform to peaked field strength across the flux rope.

Next, exploring non-circular cross sections, we found that the aspect ratio, $b$, of the cross section is the main parameter affecting the estimated impact parameter $p$.   
For flux ropes flatter in the propagating direction (corresponding to $b>1$),  $p$ is more biased to lower values, compared to the true one, as $b$ is increased.  This effect implies simulated distributions $\pp (p)$ which are close to observed ones with $b\approx 2$ for an elliptical model with uniform axial current density.  For linear force-free fields with elliptical cross-section, $p$ is less affected by $b$, but the field strength decreases more rapidly away from the flux-rope axis, so that the selection effect on the field strength enhances the dependence of $\pp (p)$ on $b$.

We also explore other effects which can bias the probability distribution of $p$.  We found that bending the cross section in a bean-like shape has a small effect on the estimated $p$. A much larger effect is present if the cross section is set broader than an ellipse at large distance from the axis.   An extreme case is a rectangular cross section.  In that case, the linear force free model corresponds to an even more biased $p$ than the above      
elliptical model with uniform axial current density, and $b \approx 1.5$ is sufficient to reproduce the observed distribution $\pobsp (p)$.   Finally, we found that for all the models explored, the rotation angle along the spacecraft trajectory is above $90\degree$ except for large $p$ values (at least $p \ge 0.7$).
Then, a selection effect on this parameter cannot explain $\pobsp (p)$.  Furthermore, only a selection criterium around $40\degree$ can lead to a computed $\pp (p)$ in agreement with $\pobsp (p)$ for large $p$ values.   This is in agreement with the minimum rotation angle found in the set of MCs analyzed by \citet{Lepping10}.

We conclude that the observed distribution $\pobsp (p)$ is mainly shaped by the oblatness of the MC cross section, with some contribution of a field strength selection when the flux rope is close to a linear force-free field.  Still, even in this last case, typically more than 70\% of the flux ropes are expected to be detected.   Even adding unfrequent cases not detected because of very large perturbations (so that the field rotation is not detected), or with a crossing within a leg, or MCs strongly in interactions, this implies a low amount of undetected flux rope, well below 2/3.   So we conclude that a large majority of flux ropes are expected to be detected. 
The non-MC ICMEs are either encountered outside the flux rope limits or they would contain none. 

  We also get results beyond the initial questions.  The main dependence of $\pp (p)$ on the aspect ratio $b$ allows to constrain a key property of the distribution $\pb (b)$ for MCs: the mean of the aspect ratio.  With an elliptical model with uniform current density, $\pobsp (p)$ sets the mean of $b$ nearby $2.3$ independently of the broadness of the distribution. 
This last property is approximately kept for a linear-force field,  but the mean of $b$ is shifted to around $3$ with a slight dependence on the amount of the selection effect of the field strength.  Then, we conclude that the observed $\pobsp (p)$ implies that MCs are moderately oblate at 1~AU, at least in average.

We further analyze the observed MCs by separating them in groups with different physical parameters.   In contrast to most MCs, the faster MCs (above $\approx 550$ km/s) have a flat $\pobsp (p)$ distribution.  It implies that the faster MCs, which typically have also both larger radius and field strength, are nearly round, while the slower ones have typically the above mean oblateness.  
Finally, we find two results indicating that the typical magnetic field profile in MCs is in between a linear-force-free field and one with a constant axial current density: 
 
- First, the mean field strength observed along the spacecraft trajectory is systematically above what is predicted by a linear force-free field,  
but below the prediction given by a constant current model, 
and this is independent of the aspect ratio of the cross section, in agreement with a previous study \citep{Gulisano05}. 

- Second, the distribution $\pp (p)$ computed with a distribution of $b$, derived to fit $\pobsp (p)$, shows systematic biases, both at low and large $p$ values, with opposite tendency for both type of magnetic fields. This conclusion is also coherent with the flatter field strength profile found in MCs compared to a Lundquist field.  Then, both the current distribution and the oblateness of the flux ropes contribute to a relatively flat profile of the field strength.

\begin{acknowledgements}
The authors thank the referee for reading carefully, and so, improving the manuscript.
The authors acknowledge financial support from ECOS-Sud
through their cooperative science program (N$^o$ A08U01).
This work was partially supported by the Argentinean grants:
UBACyT 20020090100264, PIP 11220090100825/10 (CONICET), and PICT-2007-856
(ANPCyT).
S.D. is member of the Carrera del Investigador Cien\-t\'\i fi\-co, CONICET.
S.D. acknowledges support from the Abdus Salam International Centre
for Theoretical Physics (ICTP), as provided in the frame of his
regular associateship.
The work of M.J. is funded by a contract from the AXA Research Fund.
\end{acknowledgements}
 
\bibliographystyle{aa}
\bibliography{mc}
\IfFileExists{\jobname.bbl}{}
{\typeout{}
\typeout{****************************************************}
\typeout{****************************************************}
\typeout{** Please run "bibtex \jobname" to optain}
\typeout{** the bibliography and then re-run LaTeX}
\typeout{** twice to fix the references!}
\typeout{****************************************************}
\typeout{****************************************************}
\typeout{}
}

\end{document}